\documentclass[11pt,a4paper]{article}
\usepackage{jcappub}
\usepackage{slashed}
\usepackage{braket}
\usepackage{hyperref}
\usepackage{float}
\usepackage{amssymb}
\usepackage{mathrsfs}
\usepackage{afterpage}

\usepackage[normalem]{ulem}

\newcommand{\Aa}[2]{A_{\alpha,#1}^{#2}}
\newcommand{\ain}{a_\mathrm{in}}

\newcommand{\classcode}{{\tt CLASS}}
\newcommand{\cosmicenu}{{\tt Cosmic-Enu}}
\newcommand{\DDnu}{\Delta^2_{\nu}}
\newcommand{\delcb}{\delta_\mathrm{cb}}

\newcommand{\delK}{\delta^\mathrm{(K)}}
\newcommand{\delmono}{\delta_{\ell=0}}
\newcommand{\dhdm}{\delta_{\mathrm{\HDM}}}

\newcommand{\Dmmthree}{\Delta m^2_{31}}
\newcommand{\Dmmtwo}{\Delta m^2_{21}}

\newcommand{\EHDM}{{\scriptscriptstyle EHDM}}
\newcommand{\EQ}[1]{Eq.~(\ref{#1})}
\newcommand{\EQS}[2]{Eqs.~(\ref{#1}-\ref{#2})}

\newcommand{\Fehdm}{F_{\rm \EHDM}}
\newcommand{\Ffd}{F_{\rm FD}}
\newcommand{\FFTM}{{\tt FlowsForTheMasses}}
\newcommand{\FFTMII}{{\tt FlowsForTheMasses-II}}

\newcommand{\Fhdms}{F_{\rm \HDM}^{(s)}}

\newcommand{\fnu}{f_\nu}

\newcommand{\ghdms}{g_{\rm \HDM}^{(s)}}
\newcommand{\Hc}{{\mathcal H}}
\newcommand{\Hco}{{\mathcal H}_0}
\newcommand{\HDM}{{\scriptscriptstyle HDM}}
\newcommand{\hyphi}{{\tt hyphi}}
\newcommand{\Ia}[2]{I_{\alpha, #1}^{#2}}
\newcommand{\Ifd}{I^{\rm (FD)}}
\newcommand{\intq}{\int_{\vec q}}
\newcommand{\Isk}{I^{\rm (sk)}}
\newcommand{\kfs}{k_\mathrm{FS}}
\newcommand{\kfsa}{k_{\mathrm{FS},\alpha}}
\newcommand{\kThr}{k_\mathrm{st}}
\newcommand{\mehdm}{m_{\rm \EHDM}}
\newcommand{\MFLR}{{\tt MuFLR}}
\newcommand{\mhdm}{m_{\rm \HDM}}
\newcommand{\mhdms}{m_{\rm \HDM}^{(s)}}

\newcommand{\Mnu}{M_\nu}
\newcommand{\Mnucool}{M_\nu^{\rm (cool)}}
\newcommand{\Mnuskew}{M_\nu^{\rm (skew)}}
\newcommand{\MuFLR}{{\tt MuFLR}}
\newcommand{\Neff}{N_{\rm eff}}
\newcommand{\Nglq}{N_{\rm GLQ}}
\newcommand{\Nhdm}{N_{\rm \HDM}}

\newcommand{\Nmu}{N_\mu}
\newcommand{\Nmuai}{N_{\mu,AI}}
\newcommand{\Nmunl}{N_{\mu,\mathrm{NL}}}
\newcommand{\nskew}{n_{\rm sk}}
\newcommand{\Ntau}{N_\tau}
\newcommand{\Oa}{\Omega_\alpha}
\newcommand{\Oao}{\Omega_{\alpha,0}}

\newcommand{\Obo}{\Omega_{\mathrm{b},0}}
\newcommand{\Ocb}{\Omega_\mathrm{cb}}
\newcommand{\Ocbo}{\Omega_{\mathrm{cb},0}}

\newcommand{\Om}{\Omega_\mathrm{m}}

\newcommand{\Omo}{\Omega_{\mathrm{m},0}}

\newcommand{\Ono}{\Omega_{\nu,0}}
\newcommand{\Pa}[2]{P_{\alpha,#1}^{#2}}
\newcommand{\PLeg}{{\mathcal P}}

\newcommand{\tconf}{{\mathcal T}}
\newcommand{\Tehdm}{T_{{\rm \EHDM},0}}

\newcommand{\Thdms}{T_{{\rm \HDM},0}^{(s)}}

\newcommand{\Tnu}{T_{\nu,0}}

\newcommand{\Xia}[2]{\Xi_{\alpha,#1}^{#2}}
\newcommand{\Xiatil}[2]{{\tilde\Xi}_{\alpha,#1}^{#2}}

\title{Everything hot everywhere all at once:\\Neutrinos and hot dark matter as a single effective species}
\author[a,b]{Amol Upadhye,}
\author[c,d]{Markus R. Mosbech,}
\author[b]{Giovanni Pierobon,}
\author[b]{Yvonne Y.~Y.~Wong}

\affiliation[a]{South-Western Institute for Astronomy Research,
  Yunnan University,
  Kunming 650500, People's Republic of China}
\affiliation[b]{Sydney Consortium for Particle Physics and Cosmology,
School of Physics, The University of New South Wales, 
Sydney NSW 2052, Australia}
\affiliation[c]{Institute for Theoretical Particle Physics and Cosmology (TTK),
  RWTH Aachen University, D-52056 Aachen, Germany}
\affiliation[d]{Institute for Theoretical Particle Physics (TTP), Karlsruhe Institute of Technology (KIT), 76128 Karlsruhe, Germany}

\emailAdd{a.upadhye@ynu.edu.cn}
\emailAdd{mosbech@physik.rwth-aachen.de}
\emailAdd{g.pierobon@unsw.edu.au}
\emailAdd{yvonne.y.wong@unsw.edu.au}

\abstract{
  Observational cosmology is rapidly closing in on a measurement of the sum $\Mnu$ of neutrino masses, at least in the simplest cosmologies, while opening the door to probes of non-standard hot dark matter (HDM) models.  By extending the method of effective distributions, we show that any collection of HDM species, with arbitrary masses, temperatures, and distribution functions, including massive neutrinos, may be represented as a single effective HDM species.  Implementing this method in the \FFTM{} non-linear perturbation theory for free-streaming particles, we study non-standard HDM models that contain thermal QCD axions or generic bosons in addition to standard neutrinos, as well as non-standard neutrino models wherein either the distribution function of the neutrinos or their temperature is changed.  Along the way, we substantially improve the accuracy of this perturbation theory at low masses, bringing it into agreement with the high-resolution TianNu neutrino N-body simulation to $\approx 2\%$ at $k=0.1~h/$Mpc and to $\leq 21\%$ over the range $k \leq 1~h/$Mpc.   We accurately reproduce the results of simulations including axions and neutrinos of multiple masses.  Studying the differences between the normal, inverted, and degenerate neutrino mass orderings on their non-linear power, we quantify the error in the common approximation of degenerate masses.  We release our code publicly at {\tt http://github.com/upadhye/FlowsForTheMassesII}~.
}

\begin{document}
\begin{flushright}
	{\large \tt CPPC-2024-07}\\
        {\large \tt TTK-24-36}\\
        {\large \tt TTP24-37}
\end{flushright}

\maketitle

\section{Introduction}
\label{sec:introduction}

Cosmology is well on the way to measuring the sum $\Mnu$ of neutrino masses, a fundamental particle physics parameter, for the first time.  The cosmological upper bounds $\Mnu \leq 120$~meV from joint constraints by the Planck survey~\cite{Planck:2018vyg} and $\Mnu \leq 72$~meV by the DESI survey~\cite{DESI:2024mwx} are rapidly converging upon the lower bound of $59$~meV from terrestrial experiments~\cite{Capozzi:2017ipn,Gariazzo:2018pei,DeSalas:2018rby,Esteban:2020cvm}. However, these rely upon restrictive assumptions about the dark energy responsible for the cosmic acceleration, when allowing the dark energy equation of state and its derivative to vary weakens the $\Mnu$ bound by a factor of about $2-3$~\cite{Upadhye:2017hdl,Shao:2024mag,RoyChoudhury:2024wri}.  Persistent tensions and anomalies in the cosmological data, including the Hubble tension~\cite{Freedman:2024eph,Pedrotti:2024kpn,Riess:2024ohe,Khalife:2023qbu}, the $S_8$ tension~\cite{Leauthaud:2016jdb,Poulin:2022sgp,Amon:2022azi,McCarthy:2023ism}, and the CMB lensing anomaly~\cite{Planck:2018lbu,Planck:2018vyg}, hint at a more complicated dark-sector phenomenology.  Moreover, a recent DESI+Planck analysis even prefers negative $\Mnu$~\cite{Craig:2024tky,Green:2024xbb,Elbers:2024sha}, an unphysical result possibly arising from a combination of the DESI preference for a slightly higher Hubble parameter $H_0$ and the Planck preference for unexpectedly strong lensing of the CMB.  

In this context, it is imperative that theoretical cosmologists study a broad range of phenomena associated with neutrinos and other hot dark matter (HDM) models.  Within the neutrino sector,  differences among the masses of the three species will be amplified by non-linear clustering at small scales, while non-standard models could modify the number, temperature, or distribution function of cosmological neutrinos~\cite{Oldengott:2019lke,Escudero:2022gez,MicroBooNE:2022sdp,Balantekin:2023jlg}.  Other HDM models include the axion, theorized as a solution to the strong CP problem in quantum chromodynamics~\cite{Peccei:1977hh,Peccei:1977ur},  whose thermal production in the early universe in the $m\lesssim$ eV regime was recently studied in Refs.~\cite{Hannestad:2010yi,Archidiacono:2013cha,Ferreira:2018vjj,DEramo:2021psx,DEramo:2021lgb,Notari:2022ffe,DEramo:2022nvb,Bianchini:2023ubu}. Each of these HDM models modify the gravitational clustering of HDM, leading to subtle differences in their clustering power as well as their impact upon scale-dependent halo bias~\cite{LoVerde:2014pxa,Chiang:2018laa}, differential HDM capture by halos between HDM-rich and HDM-poor regions~\cite{Yu:2016yfe}, ``wakes'' caused by coherent HDM streaming past collapsed cold dark matter (CDM) halos~\cite{Zhu:2014qma}, and other non-linear effects.  As cosmological data improve, such higher-order effects could be used either to confirm the standard neutrino picture or to reject it in favor of a more complicated HDM sector, provided that we have the theoretical and numerical tools with which to quantify HDM clustering. Since the space of HDM models is large, tools such as non-linear perturbation theory allowing for a rapid exploration of the parameter space are desirable.

However, hot dark matter presents a host of numerical challenges due to its large velocity dispersion.  A given HDM particle's thermal velocity acts as an escape velocity allowing it to stream freely out of a sufficiently small and diffuse halo. Whereas the CDM and baryons are cold, in the sense that their velocities are determined entirely by their positions, flattening their six-dimensional phase space to a three-dimensional sheet, we must track all six dimensions for an HDM species.   In an N-body particle simulation, sampling this six-dimensional phase space requires a large number of particles to avoid shot noise, while faster particles require a finer time stepping to track their gravitational deflection.  Velocity dispersion also complicates a perturbative treatment of HDM, since non-linear perturbation theories typically begin with the continuity and Euler equations, which assume a well-defined velocity field.  These challenges are compounded by the need to include standard neutrinos, at least two of which are massive, along with any additional HDM species.  Each of these HDM species has its own mass and momentum distribution.

We make two contributions to this velocity dispersion problem.  
Firstly, we extend the effective distribution function method of Ref.~\cite{Bayer:2020tko}, originally developed for non-relativistic neutrinos, to represent any collection of HDM as a single \emph{effective hot dark matter} (EHDM). Our extension applies to relativistic as well as non-relativistic species of different masses, temperatures, and distribution functions, provided that each species has decoupled from all non-gravitational interactions.  Our key insight is that the clustering of such a particle depends only upon its four-velocity, rather than its mass and four-momentum separately, so that doubling both an HDM particle's mass and its momentum simultaneously will not affect its clustering.  

We implement the EHDM method in the \FFTM{} perturbation theory of Ref.~\cite{Chen:2022cgw}, which provided the first non-linear perturbative power spectrum computation for free-streaming HDM.  \FFTM{} functions by discretizing the HDM distribution function into ``flows,'' each characterized by its four-velocity, so that a single EHDM flow can represent many different HDM species.  Furthermore, we show that the clustering power of an individual HDM component of the EHDM can be recovered from a linear combination of these same EHDM flows.  This is crucial, for example, for a terrestrial detector that is sensitive only to the electron neutrino, or only to the axion, rather than to all HDM species making up the EHDM.

Secondly, we trade the uniform-density momentum binning of Ref.~\cite{Chen:2022cgw} for a more efficient binning based upon Gauss-Laguerre quadrature, taking advantage of the fact that thermal distribution functions decay exponentially with particle momentum.  Implementing our new procedure in the code \FFTMII{}, we demonstrate the effectiveness of this improved quadrature by reducing the low-$\Mnu$, high-$k$ error found in Ref.~\cite{Upadhye:2023bgx} by more than a factor of two.  We demonstrate the accuracy of \FFTMII{} for models containing axions and axion-like bosons in addition to neutrinos, and we apply it to quantifying the error in the commonly-used approximation treating all three neutrino masses as equal.  Finally, we consider a pair of proposals to evade cosmological bounds on high-mass neutrinos, one by raising and the other by lowering their mean momentum~\cite{Oldengott:2019lke,Escudero:2022gez}.  While the high-mean-momentum neutrinos are nearly indistinguishable from standard ones, the low-mean-momentum ones cluster more non-linearly, in a manner that will allow upcoming cosmological surveys to search for them.  Thus we demonstrate the ability of non-linear perturbation theory to explore a large parameter space of non-standard models.

This article is organized as follows.  After summarizing the numerical techniques used, in Sec.~\ref{sec:bkg}, we derive and thoroughly study the EHDM technique in Sec.~\ref{sec:eff}, and describe its implementation along with Gauss-Laguerre quadrature in \FFTMII{}.  Our results are split into two sections, beginning with Sec.~\ref{sec:re1}, which quantifies the non-linear enhancement of HDM clustering in a range of models.  Section~\ref{sec:re2} studies the two proposals for evading cosmological $\Mnu$ bounds mentioned above, and Sec.~\ref{sec:con} concludes.

\section{Background}
\label{sec:bkg}

\subsection{Cosmic neutrinos}
\label{subsec:bkg:cosmic_neutrinos}

The Standard Model of particle physics predicts exactly three neutrinos, which are uncharged, weakly-interacting fermions whose small masses make them ultrarelativistic for much of the universe's history up to CMB formation.  Neutrinos decouple from photons, electrons, and positrons at a temperature of $T \sim 1$~MeV, shortly before electron-positron annihilation begins.  In an idealized situation, the neutrino temperature at the end of $e^+ e^-$ annihilation is $(4/11)^{1/3}$ times that of the photons. Detailed calculations, however, have found that the neutrino-to-photon energy density is some $1.47\%$ larger than that suggested by this simple temperature relation.  This is equivalent to an increase in the effective number of neutrinos to $3.044$~\cite{Bennett:2019ewm,Akita:2020szl,Froustey:2020mcq,Bennett:2020zkv,Drewes:2024wbw}, which we approximate by raising the neutrino temperature by $0.365\%$, to $\Tnu a^{-1} = 1.9525 a^{-1}$~K at scale factor $a$.

Much later, typically around or after electron-photon decoupling, the behavior of cosmic neutrinos is determined by their masses, at least two of which are required to be non-zero by neutrino oscillation experiments.  Such experiments can determine only mass-squared splittings $\Delta m_{21}^2 = 74.2^{+2.1}_{-2.0}$~meV$^2$ and $|\Delta m_{31}^2| = 2517^{+26}_{-28}$~meV$^2$ rather than the absolute neutrino mass scale $\Mnu$~\cite{Esteban:2020cvm}.  Furthermore, the sign of the larger splitting may be either positive or negative, with the former implying a ``normal order'' (NO) of neutrino masses dominated by a single heavy neutrino and two light ones, and the latter implying an ``inverted order'' (IO) with two heavy neutrinos and a single light one.  As $\Mnu$ rises, the fractional difference between the heaviest and lightest decreases, making a ``degenerate order'' (DO) of three equal neutrino masses a common approximation. 

Even at late times, neutrinos' Fermi-Dirac thermal velocity distribution profoundly affects their clustering.  The subhorizon, non-relativistic clustering of a neutrino species of mass $m_\nu$ is characterized by a ``free-streaming'' length that is approximately the average distance travelled by neutrinos of that mass in a comoving Hubble time ${\cal H}^{-1}$.  On much larger length scales, neutrinos' free-streaming does not inhibit their clustering, and they cluster much the same as cold matter; we call this the ``clustering regime.''  On scales smaller than the free-streaming length, the ``free-streaming regime,'' neutrinos stream out of most overdense regions.  We define the neutrino free-streaming wave number as~\cite{Ringwald:2004np}
\begin{equation}
  \kfs := \sqrt{ \frac{3\Om(a) \Hc(a)^2}{2 c_\nu(a)^2}},
\end{equation}
where the square of the neutrino sound speed is 
\begin{equation}  
  c_\nu(a)^2 := \frac{3\zeta(3) \Tnu^2 }{2\ln(2) m_\nu^2 a^2},
  \label{e:def_kfs_cnu}
\end{equation}
and $\zeta(x)$ is the Riemann zeta function.
Refs.~\cite{Ringwald:2004np,Wong:2008ws} demonstrated that the linear perturbation ratio $\delta\rho_\nu / \delta\rho_{\rm m} / (\bar\rho_\nu / \bar\rho_{\rm m})$ approaches unity at small $k$ and $\kfs^2/k^2$ at large $k$, leading them to approximate
\begin{equation}
  \frac{\delta\rho_\nu(k)}{\delta\rho_{\rm m}(k)}
  \approx
   \frac{\bar\rho_\nu}{\bar\rho_{\rm m}} \frac{1}{(1 + k/\kfs)^2},
  \label{e:super-easy_approx}
\end{equation}
by interpolating between the clustering and free-streaming limits.

\subsection{Multi-fluid perturbation theories}
\label{subsec:bkg:multi-fluid_perturbation_theories}

The chief difficulty with applying standard cosmological perturbation theory to HDM species such as massive neutrinos is their significant velocity dispersion.  Whereas a cold particle beginning at a given initial position with zero velocity can be tracked to a definite final position, HDM particles begin with a thermal distribution of initial velocities.  Thus we must track their full six-dimensional phase space distribution, rather than the three-dimensional spatial distribution of cold particles.

In a series of articles, Dupuy and Bernardeau demonstrated that neutrinos could be treated perturbatively by splitting their population into multiple sub-populations, each characterized by a spatially-uniform zeroth-order velocity $\vec v$~\cite{Dupuy:2013jaa,Dupuy:2014vea,Dupuy:2015ega}.  Since a particle with definite initial velocity can once again be tracked from a given initial position to a definite final position, standard perturbative techniques may be applied.  Furthermore, the direction of $\vec v$ affects clustering only through its angle with the Fourier vector.  Thus the multi-fluid method increases the dimensionality of the problem by two, $v$ and $\hat v \cdot \hat k$, rather than three.  References~\cite{Dupuy:2013jaa,Dupuy:2014vea} derived a fully relativistic linear theory for massive neutrinos in Newtonian gauge and a general gauge, respectively, while Ref.~\cite{Dupuy:2015ega} motivated a non-linear treatment; all of these results generalize to other HDM.

The EHDM formalism to be presented in Sec.~\ref{sec:eff} is applicable to the relativistic perturbation theory of Refs.~\cite{Dupuy:2013jaa,Dupuy:2014vea}.  However, our focus here is clustering at late times, when each HDM species is non-relativistic, and particularly on small-scale non-linear HDM clustering.  Thus we focus on multi-fluid perturbation theories in the subhorizon, non-relativistic case.  In this limit, we may to excellent approximation treat fluids as obeying the continuity and Euler equations of classical fluid dynamics, with a gravitational potential determined from Poisson's equation, in a universe whose uniform expansion is given by the Hubble rate.  We restrict our consideration to spatially-flat cosmologies, though our results may be generalized to models with spatial curvature.  Finally, we track only the scalar perturbations of each fluid, namely, the density contrast and velocity divergence.  Although vector perturbations such as the velocity vorticity are important in small-scale, non-perturbative structures such as virialized halos, they are negligible at the linear and mildly non-linear scales accessible to perturbation theory~\cite{Pueblas:2008uv,Jelic-Cizmek:2018gdp,Umeh:2023lbc}. 

References~\cite{Dupuy:2013jaa,Dupuy:2014vea} discretized the Fermi-Dirac distribution describing the initial neutrino momenta.  Let $P_i$ be one such lower-index three-momentum, and let $\tau_i$ be its value in the limit of a homogeneous universe.  Then $P_i \rightarrow \tau_i$ at early times, and $\tau_i$ is itself time-independent, making it a useful quantity for a Lagrangian description of neutrinos in momentum space.  Furthermore, physics depends upon the direction of $\vec \tau$ only through its angle $\mu = \cos^{-1}(\hat \tau \cdot \hat k)$ with the Fourier vector $\vec k$, so we may approximate the neutrino population using $\Ntau$ values of the momentum magnitude $\tau = \left| \vec\tau \right| = \sqrt{\tau_1^2 + \tau_2^2 + \tau_3^2}$.  We label these discrete momenta using Greek indices, as $\tau_\alpha$, with integer $\alpha \in [0,\Ntau-1]$.  

Reference~\cite{Chen:2020bdf} restricted the perturbation theory of Refs.~\cite{Dupuy:2013jaa,Dupuy:2014vea} to the subhorizon, non-relativististic case, which is  released as the code \MFLR{}.%
\footnote{\MFLR{} is publicly available at {\tt github.com/aupadhye/MuFLR}~.} %
The scalar perturbations of each ``flow'' $\alpha$ are the density contrast $\delta_\alpha(\vec x) := \rho_\alpha(\vec x) / \bar\rho_\alpha - 1$ and the velocity divergence $\theta_\alpha(\vec x) := -\vec\nabla \cdot \vec P / (\mhdm a)$, where $\mhdm$ is the HDM mass.%
  \footnote{Strictly speaking, the continuity and Euler equations apply to a fluid with a well-defined momentum, rather than merely a momentum magnitude, at each point in space.  However,  all fluids with the same momentum magnitude obey the same set of equations of motion.  We use the term ``flow'' to refer to all such fluids at once, and the same index $\alpha$ for all fluids with initial momentum magnitude $\tau_\alpha$.} %
  In Fourier space, $\delta_\alpha$ and $\theta_\alpha$ depend upon the magnitude $k$ of the wave number as well as its angle with respect to the initial momentum $\vec\tau_\alpha$, through its cosine $\mu := \hat k \cdot \hat \tau$. The $\mu$-dependence of these perturbations may be expanded in Legendre polynomials as
  \begin{equation}
    \delta_\alpha^{\vec k}
    := \sum_{\ell=0}^\infty (-i)^\ell \PLeg_\ell(\mu) \delta^k_{\alpha\ell},
    \quad\quad \quad
    \theta_\alpha^{\vec k}
    := \sum_{\ell=0}^\infty (-i)^\ell \PLeg_\ell(\mu) \theta^k_{\alpha\ell},
  \end{equation}
  where $\PLeg_\ell$ is the Legendre polynomial of order $\ell$; we use wave number superscripts to denote functional dependence, so $\delta_\alpha^{\vec k} = \delta_\alpha(\vec k)$.

  Using $\eta := \log(a/\ain)$ as our time variable,\footnote{Throughout this article, we use log to denote the natural logarithm.} for initial scale factor $\ain$, and primes to denote derivatives with respect to $\eta$, the linear continuity and Euler equations for flow $\alpha$ are
\begin{equation}
\begin{aligned}
  (\delta^k_{\alpha\ell})'
  &=
  \frac{kv_\alpha}{\Hc}
  \left( \frac{\ell}{2\ell-1} \delta^k_{\alpha,\ell-1}
  - \frac{\ell+1}{2\ell+3} \delta^k_{\alpha,\ell+1}\right)
  + \theta^k_{\alpha\ell},
  \label{e:eom_delta_lin}
  \\
  (\theta^k_{\alpha\ell})'
  &=
  -\left(1+\frac{\Hc'}{\Hc}\right)\theta^k_{\alpha\ell}
  - \delK_{\ell 0} \frac{k^2\Phi^k}{\Hc^2}
  + \frac{kv_\alpha}{\Hc}
  \left( \frac{\ell}{2\ell-1} \theta^k_{\alpha,\ell-1}
  - \frac{\ell+1}{2\ell+3} \theta^k_{\alpha,\ell+1}\right),
  \end{aligned}
\end{equation}
where $\delK$ is the Kronecker delta, $\Hc = aH$  the conformal Hubble rate, and $v_\alpha := \tau_\alpha/(\mhdm a)$  the flow speed.  Since this perturbation theory is Lagrangian in momentum space, a particle cannot move from one flow to another.  Thus the only interaction between different flows occurs through the gravitational potential $\Phi$, given by Poisson's equation:
\begin{equation}
  k^2 \Phi^k
  =
  -\frac{3}{2} \Hc^2
  \left( \Ocb(\eta) \delcb^k
  + \sum_{\alpha=0}^{\Ntau-1} \Oa(\eta) \delta^k_{\alpha 0} \right).
  \label{e:eom_Poisson}
\end{equation}
Here, $\delcb$ is the density contrast of the CDM and baryons, which we approximate as a single fluid labeled ``cb'' henceforth.  It can be obtained either from a linear perturbative treatment of the cb fluid, resulting in a fully linear perturbation theory for the neutrinos, or from a non-linear calculation which then sources the linear \EQ{e:eom_delta_lin} through \EQ{e:eom_Poisson}, known as ``linear response.''  The time-dependent density fractions $\Ocb(\eta)$ and $\Oa(\eta)$ are respectively given in terms of their values $\Ocbo$ and $\Oao$ today by $\Hc^2 \Ocb(\eta) = \Hco^2 \Ocbo / a$ and $\Hc^2 \Oa(\eta) = \Hco^2 \Oao / (a \sqrt{1-v_\alpha^2})$, where $\Hco$ is the value of $\Hc$ today.

Initial conditions at $\ain=10^{-3}$ are given in Ref.~\cite{Chen:2020bdf}.  Specifically,
\begin{equation}
  \delcb(\ain) = \ain + \frac{2}{3}a_{\rm eq},
  \quad\quad\quad
  \theta_{\rm cb}(\ain) = \ain,
  \label{e:ICs_delcb_thetacb}
\end{equation} 
and  
\begin{equation}
  \delta_{\alpha,0}(\ain,k)
  =
  \frac{\kfsa^2 (1-\fnu) \delcb(\ain)}{(k+\kfsa)^2 - \fnu\kfsa^2},
  \quad\quad\quad
  \theta_{\alpha,0} = \delta_{\alpha,0}',
  \label{e:ICs_delhdm_thetahdm}
\end{equation}
for cb and HDM, respectively, where
\begin{equation}
  \kfsa^2
  =
  \frac{3 \Om(a) \Hc^2 a^2 \mhdm^2}{2\tau_\alpha^2}
  \label{e:kfs_alpha}
\end{equation}
is a generalization of \EQ{e:def_kfs_cnu} to a flow of arbitrary velocity $v_\alpha = \tau_\alpha/(\mhdm a)$, shown in Ref.~\cite{Chen:2020bdf} to obey $\delta_{\alpha,0}/\delta_{\rm m} \rightarrow \kfsa^2/k^2$ at large $k$.  The HDM initial density monopole of \EQ{e:ICs_delhdm_thetahdm} is an interpolation between the clustering limit, $\delta_{\alpha,0} \approx \delcb$, and the free-streaming limit, $\delta_{\alpha,0} \approx (k/\kfsa)^2 (1-\fnu)\delcb$, similar to that of \EQ{e:super-easy_approx}, from Refs.~\cite{Ringwald:2004np,Wong:2008ws,Pierobon:2024XX}.

Reference~\cite{Chen:2022cgw} developed the first non-linear perturbative power spectrum calculation for free-streaming HDM, such as massive neutrinos, called \FFTM{}.%
\footnote{\FFTM{} is publicly available at {\tt github.com/upadhye/FlowsForTheMasses}~.} %
It did so by generalizing the Time-Renormalization Group (Time-RG) perturbation theory of Refs.~\cite{Pietroni:2008jx,Lesgourgues:2009am} to the case of a fluid with zeroth-order bulk velocity $\vec v_\alpha$.  The result is a set of perturbative mode-coupling integrals that couple different multipoles $\ell$ as well as wave numbers~$k$.  Since non-linear corrections decorrelate $\delta_\alpha$ and $\theta_\alpha$, Ref.~\cite{Chen:2022cgw} introduced the decorrelation perturbation
\begin{equation}
  \chi^k_{\alpha\ell}
  :=
  1  - \Pa{01\ell}{k} / \sqrt{\Pa{00\ell}{k} \Pa{11\ell}{k}},
\end{equation}
where $\Pa{bc\ell}{k}$ is the $\ell$th Legendre moment of the power spectrum,
\begin{equation}
\begin{aligned} 
  \Pa{00}{\vec k}
  =& \sum_\ell \PLeg_\ell(\mu)^2 \Pa{00\ell}{k},
  \quad
  \Pa{11}{\vec k}
  = \sum_\ell \PLeg_\ell(\mu)^2 \Pa{11\ell}{k},
  \\
  \Pa{01}{\vec k}
  =&
  \sum_\ell \PLeg_\ell(\mu)^2 (1-\chi^k_{\alpha\ell})
  \sqrt{\Pa{00\ell}{k} \Pa{11\ell}{k}},
  \end{aligned}
\end{equation}
and its indices $0$ and $1$ refer respectively to $\delta$ and $\theta$.  The non-linear equations of motion of \FFTM{}, which replace \EQ{e:eom_delta_lin}, are then
\begin{equation}
\begin{aligned}
  (\delta^k_{\alpha\ell})'
  =&
  \tfrac{kv_\alpha}{\Hc}
  \left( \tfrac{\ell}{2\ell-1} \delta^k_{\alpha,\ell-1}
  - \tfrac{\ell+1}{2\ell+3} \delta^k_{\alpha,\ell+1}\right)
  + \theta^k_{\alpha\ell}
  + \tfrac{2}{\delta^k_{\alpha\ell}} \Ia{001,001,\ell}{k}\, ,
  \\
  (\theta^k_{\alpha\ell})'
  =&
  -\left(1+\tfrac{\Hc'}{\Hc}\right)\theta^k_{\alpha\ell}
  - \delK_{\ell 0} \tfrac{k^2\Phi^k}{\Hc^2}
  + \tfrac{kv_\alpha}{\Hc}
  \left( \tfrac{\ell}{2\ell-1} \theta^k_{\alpha,\ell-1}
  - \tfrac{\ell+1}{2\ell+3} \theta^k_{\alpha,\ell+1}\right)
  + \tfrac{1}{\theta^k_{\alpha\ell}} \Ia{111,111,\ell}{k}\, ,
  \\
  (\chi^k_{\alpha\ell})'
  =&
  \tfrac{2(1-\chi^k_{\alpha\ell})}{(\delta^k_{\alpha\ell})^2} \Ia{001,001,\ell}{k}
  + \tfrac{1-\chi^k_{\alpha\ell}}{(\theta^k_{\alpha\ell})^2} \Ia{111,111,\ell}{k}
  - \tfrac{2}{\delta^k_{\alpha\ell}\theta^k_{\alpha\ell}} \Ia{001,101,\ell}{k}
  - \tfrac{1}{\delta^k_{\alpha\ell}\theta^k_{\alpha\ell}} \Ia{111,011,\ell}{k}\, .
  \label{e:eom_chi_nl}
  \end{aligned}
\end{equation}
Here, the bispectrum integrals $\Ia{acd,bef,\ell}{k}$ are defined by their equations of motion,
\begin{equation}
  (\Ia{acd,bef,\ell}{k})'
  =
  - \Xia{bg\ell}{k} \Ia{acd,gef,\ell}{k}
  - \Xiatil{eg\ell}{k} \Ia{acd,bgf,\ell}{k}
  - \Xiatil{fg\ell}{k} \Ia{acd,beg,\ell}{k}
  + 2\Aa{acd,bef,\ell}{k}\, ,
  \label{e:eom_I_nl}
\end{equation}
with
\begin{equation}
\begin{aligned}
  \Xia{bc\ell}{k}
  =&
  \left[ \begin{array}{cc}
      0                                          & -1                   \\
      \tfrac{k^2 \Phi^i}{\Hc^2 \delta^k_{\alpha 0}} & 1 + \tfrac{\Hc'}{\Hc}
    \end{array} \right]
  - \tfrac{k v_\alpha}{\Hc}\tfrac{\ell}{2\ell-1}
    \left[ \begin{array}{cc}
        \tfrac{\delta^k_{\alpha,\ell-1}}{\delta^k_{\alpha,\ell}} & 0  \\
        0 & \tfrac{\theta^k_{\alpha,\ell-1}}{\theta^k_{\alpha,\ell}}
    \end{array} \right]
  + \tfrac{k v_\alpha}{\Hc}\tfrac{\ell+1}{2\ell+3}
    \left[ \begin{array}{cc}
        \tfrac{\delta^k_{\alpha,\ell+1}}{\delta^k_{\alpha,\ell}} & 0  \\
        0 & \tfrac{\theta^k_{\alpha,\ell+1}}{\theta^k_{\alpha,\ell}}
      \end{array} \right],\,
  \\
  \Xiatil{bc\ell}{k}
  =&
  \left[ \begin{array}{cc}
      0 & -1                   \\
      0 & 1 + \tfrac{\Hc'}{\Hc}
    \end{array} \right],
\end{aligned}
\end{equation}
where the indices $b$ and $c$ label the rows and columns respectively, 
\begin{eqnarray}
  \Aa{acd,bef}{\vec k}
 & =&
  \intq \gamma_{acd}^{\vec k \vec q \vec p}
  \left[
    \gamma_{bgh}^{\vec k \vec q \vec p}\Pa{ge}{\vec q}\Pa{hf}{\vec p}
    + \gamma_{egh}^{\vec q, -\vec p, \vec k}\Pa{gf}{\vec p}\Pa{hb}{\vec k}
    + \gamma_{fgh}^{\vec p, \vec k, -\vec q}\Pa{gb}{\vec k}\Pa{he}{\vec q}
    \right]
  \label{e:def_A}
  \\
  &=:& \sum_\ell \PLeg_\ell(\mu)^2 \Aa{acd,bef,\ell}{k}\, 
  \label{e:def_A_ell}
\end{eqnarray}
is the mode-coupling integral, 
and
\begin{equation}
    \gamma_{001}^{\vec k \vec q \vec p}
  =
  \gamma_{010}^{\vec k \vec p \vec q} = \frac{(\vec q + \vec p)\cdot \vec p}{2p^2},
  \quad \quad
  \gamma_{111}^{\vec k \vec q \vec p}
  = \frac{|\vec q + \vec p|^2 \vec q \cdot \vec p}{2q^2 p^2},
\end{equation}
while all other $\gamma_{abc}$ vanish.
Their initial conditions, set at $\eta=0$ (i.e., $a=a_{\rm in}$), are
\begin{equation}
  \Ia{acd,bef,\ell}{k} = 2 \Aa{acd,bef,\ell}{k}.
  \label{e:eom_I_ICs}
\end{equation}
On the right hand sides of Eqs.~(\ref{e:eom_I_nl},~\ref{e:def_A}), we assume summation over repeated indices.  Computation of the mode-coupling integrals $\Aa{acd,bef,\ell}{k}$ of \EQ{e:def_A_ell} is the most expensive part of \FFTM{}, and its acceleration using Fast Fourier Transform (FFT) techniques is described thoroughly in Ref.~\cite{Chen:2022cgw}.

Thus far we have not discussed precisely how $\tau_\alpha$ are to be sampled from the Fermi-Dirac distribution function $\Ffd$, or another distribution function appropriate to other HDM species.  The \MFLR{} and \FFTM{} perturbation theories considered in Refs.~\cite{Chen:2020bdf,Chen:2022cgw} used equal-number-density bins.  That is, the range $0 \leq \tau < \infty$ was divided into $\Ntau$ intervals such that the integrals of $4\pi \tau^2 \Ffd(\tau)$ over any two intervals are equal.  For each $\alpha \in [0,\Ntau-1]$, $\tau_\alpha$ was chosen to be the median of the corresponding interval.  In Sec.~\ref{subsec:bkg:gauss-laguerre_quadrature} and Sec.~\ref{subsec:eff:discrete_p_and_GLQ} we will discuss a more efficient sampling method.

\subsection{Gauss-Laguerre quadrature}
\label{subsec:bkg:gauss-laguerre_quadrature}

Consider a function $g(x)$, defined on $[0,\infty)$.  Gauss-Laguerre quadrature (GLQ) approximates the integral of $e^{-x} g(x)$ on the semi-infinite interval using $\Nglq$ points $x_\alpha$ and weights $w_\alpha$ as
\begin{equation}
  \int_0^\infty dx\, e^{-x} g(x)
  \approx \sum_{\alpha=0}^{\Nglq-1} w_\alpha g(x_\alpha).
  \label{e:def_GLQ}
\end{equation}
Here, $\Nglq$ is a positive integer; the $x_\alpha$ are the $\Nglq$ roots of the $\Nglq$th Laguerre polynomial ${\mathcal L}_{\Nglq}(x)$; and the weights are given by 
\begin{equation}
  w_\alpha = \frac{x_\alpha}{(\Nglq+1)^2 {\mathcal L}_{\Nglq+1}(x_\alpha)^2}.
  \label{e:def_w_GLQ}
\end{equation}
If $g(x)$ is a polynomial of degree no more than $2\Nglq-1$, then \EQ{e:def_GLQ} is exact rather than approximate~\cite{Chandrasekhar_1960}.  We will find it convenient to define $G(x) = e^{-x} g(x)$, for which \EQ{e:def_GLQ} implies $\int_0^\infty dx\, G(x) \approx \sum_\alpha w_\alpha e^{x_\alpha} G(x_\alpha)$.

The error in approximation \EQ{e:def_GLQ}, given by Eq.~(25.4.45) of Ref.~\cite{Abramowitz_and_Stegun} is
\begin{equation}
  \epsilon_{\rm GLQ}
  =
  \frac{(\Nglq!)^2}{(2\Nglq)!} g^{(2\Nglq)}(x_\epsilon)
  \approx
  \sqrt{\pi\Nglq} 2^{-2\Nglq} g^{(2\Nglq)}(x_\epsilon)
  \label{e:err_GLQ}
\end{equation}
for some $0 \leq x_\epsilon < \infty$, where $g^{(2\Nglq)}$ is the $(2\Nglq)$th derivative of $g$.  The approximation in \EQ{e:err_GLQ} uses Stirling's formula for large $\Nglq$.  We will see that $g(x)$ is typically $x^2 e^x$ times a thermal distribution function.  For the Boltzmann distribution $e^{-x}$, $g(x)$ is precisely a polynomial, making \EQ{e:def_GLQ} exactly correct for $\Nglq \geq 2$.  Errors for the Bose-Einstein and Fermi-Dirac distributions are due to their difference from the Boltzmann distribution, differences whose derivatives are at most ${\mathcal O}(1)$, meaning that $\Nglq \gtrsim 10$ should be highly accurate.  However, at very small length scales, we will see that $g(x)$ is suppressed by an additional two powers of $x$, leading to larger errors.

We find points and weights for GLQ using the {\tt scipy.special.roots\_laguerre} python function.  This limits us to $\Nglq \leq 186$; above this bound, double precision numbers are inadequate for evaluating the higher-order Laguerre polynomials required for determining the weights.  We will see that $\Nglq=186$ is far larger than necessary for our applications.

\subsection{Effective distribution functions}
\label{subsec:bkg:effective_distribution_functions}

The method of effective distribution functions was introduced in Ref.~\cite{Bayer:2020tko} and applied to the case of multiple neutrinos with non-degenerate masses.  We summarize their method here before generalizing it in the next section.

Consider a neutrino species $s$ with mass $m_\nu^{(s)}$, temperature constant $T_{\nu,0}^{(s)}$, and lower-index homogeneous-universe three-momentum $\tau_i^{(s)}$, hence $(\tau^{(s)})^2 = (\tau_1^{(s)})^2 + (\tau_2^{(s)})^2 + (\tau_3^{(s)})^2$.  Its distribution function is the Fermi-Dirac distribution in the relativistic limit, $\Ffd(\tau^{(s)}) = (2\pi)^{-3} [\exp(\tau^{(s)}/T_{\nu,0}^{(s)}) + 1]^{-1}$.  The mass in a phase space volume element $d^3x d^3\tau^{(s)}$
is then $g_\nu^{(s)} m_\nu^{(s)}\Ffd(\tau^{(s)}) d^3x d^3\tau^{(s)}$, where $g_\nu^{(s)}=2$ accounts for a neutrino and an antineutrino.

Defining $\tau_i = \tau_i^{(s)} \mehdm /  m_\nu^{(s)}$ for some quantity $\mehdm$ with dimensions of mass, we may change phase space variables from $\tau_i^{(s)}$ to $\tau_i$.  The mass of multiple species in a phase space element may now be written as $\sum_s g_\nu^{(s)} m_\nu^{(s)} \Ffd(\tau  m_\nu^{(s)} /\mehdm ) (m_\nu^{(s)} /\mehdm)^3 d^3x d^3\tau$.  Thus if we define a new EHDM particle with mass $\mehdm$, momentum $\tau$, and distribution  
\begin{equation}
  \Fehdm(\tau)
  =
  \sum_s g_\nu^{(s)} \left(\frac{m_\nu^{(s)}}{\mehdm}\right)^4
  \Ffd\!\!\left(\frac{\tau  m_\nu^{(s)}}{\mehdm}\right),
  \label{e:def:Fehdm}
\end{equation}
then its mass density $\mehdm \Fehdm(\tau) d^3x d^3\tau$ equals that of all neutrino species combined.  Rather than including three different-mass neutrino species into an expensive calculation such as an N-body simulation, we may include a single particle with mass $\mehdm$ and the above distribution function.

\section{Hot dark matter as an effective particle}
\label{sec:eff}

\subsection{Effective HDM}
\label{subsec:eff:effective_hdm}

Before implementing the method of effective distribution functions, we discuss its  applicability to distribution functions that vary in time and space, as well as its limitations.  Working in conformal Newtonian gauge and using conformal time $\tconf$, the line element is
\begin{equation}
ds^2 = a^2[-(1+2\Phi)d\tconf^2 + (1-2\Psi)|\vec{dx}|^2].
\end{equation}
The collisionless Boltzmann equation for particles of positive mass $m$ and four-velocity $U_\mu$, hence four-momentum $P_\mu = m U_\mu$, along a geodesic with affine parameter $\lambda$ is
\begin{equation}
  0
  =
  U^\mu \frac{\partial F}{\partial x^\mu}
  + \frac{\partial U^i}{\partial \lambda} \frac{\partial F}{\partial U^i}
  =
  U^0 \frac{\partial F}{\partial \tconf} + U^i \frac{\partial F}{\partial x^i}
  - \Gamma^i_{\mu\nu} U^\mu U^\nu \frac{\partial F}{ \partial U_i},
\end{equation}
where we have used the geodesic equation, and $\Gamma^i_{\mu\nu}$ is the Christoffel symbol.  In particular, the evolution of the distribution function is independent of the particle mass.  For a given initial position and four-velocity, the fractional change in every such distribution function is identical.  Thus an EHDM with a distribution function of the form of Eq.~(\ref{e:def:Fehdm}), defined at a time after all HDM species have decoupled from any non-gravitational interactions, will continue to represent those species thenceforth.  This conclusion applies to all orders in perturbations of the metric and the distribution function.

The stress-energy tensor for this species of mass $m$ may also be expressed as an integral over four-velocities:
\begin{equation}
  T_{\mu\nu}
  = \int \frac{d^3 P_i}{\sqrt{-g}} \frac{P_\mu P_\nu}{P^0} F(x,U)
  = \int \frac{d^3 U_i}{\sqrt{-g}} \frac{U_\mu U_\nu}{U^0} m^4 F(x,U).
  \label{e:T_ehdm}
\end{equation}
This clarifies the $m^4$ scaling of each component of the effective distribution function of Eq.~(\ref{e:def:Fehdm}).  Since the $T_{\mu\nu}$ integral scales as the fourth power of the four-momentum and 
contributions from multiple species add linearly, it follows that replacing $P^\mu$ with $m U^\mu$ must lead to an effective distribution for all species~$s$ at any given position $x$ and four-velocity $U$ proportional to the sum of $g^{(s)} (m^{(s)})^4 F^{(s)}(x,U)$ over all $s$. 

Also evident from the above is a limitation of the effective distribution function approach.  The distribution function $F^{(s)}(x,U)$ for an individual HDM species $s$ records the phase-space number density of that species, and the sum over $s$ the total HDM phase-space number density.  However, the corresponding effective distribution function will not in general match the individual or total HDM number densities across all of phase space.  The $m^4$ mass scaling of Eq.~(\ref{e:def:Fehdm}) results in the correct $T_{\mu\nu}$, which has mass dimension four, but we cannot rely on it for other quantities.  Fortunately for our purposes, $T_{\mu\nu}$ is sufficient for studying our observables of interest as well as their evolution and their impact on the spacetime metric through Einstein's equation.

Since we are particularly interested in density perturbations, we next simplify the $T^0_0$ component of the stress-energy tensor.  Dupuy and Bernardeau point out in Ref.~\cite{Dupuy:2013jaa} that the spatial components of the lower-index four-velocity in the limit of a homogeneous universe, $U^{(0)}_i$, are constant in time.  Thus they treat the constant $\vec u := (U^{(0)}_1, U^{(0)}_2, U^{(0)}_3)^T$ as a Lagrangian coordinate for the particle velocity.  Note that $u_0 := U^{(0)}_0 = -\sqrt{a^2 + |\vec u|^2}$, where $|\vec u|^2 = \delK_{ij} u_i u_j$.  Our treatment in Sec.~\ref{subsec:bkg:multi-fluid_perturbation_theories} further simplifies the perturbation theory by working in the subhorizon, non-relativistic limit, in which the flow velocity is $\vec v \approx \vec u / a$.

Making the cosmologically valid approximation that the metric potentials $\Phi$ and $\Psi$ are small compared with unity, even though the matter perturbations may be large, we have $U_0 = (1+\Phi)U^{(0)}_0$ and $U_i = (1-\Psi)U^{(0)}_i$.  Since $\sqrt{-g} = a^4(1+\Phi-3\Psi)$, we may simplify $d^3U_i / (U^0 \sqrt{-g}) \approx -d^3 \vec u / (a^2 u_0)$ to linear order in $\Phi$ and $\Psi$.  Up to the same order of approximation, these also cancel from the products $U^0 U_0 = -a^{-2} u_0^2$ and $U^i U_j = a^{-2} u_i u_j$.

Next, let ${\bar T}_{\mu\nu}$ be the spatially-averaged stress-energy tensor, representing the homogeneous component of the matter.  Because $F(x,U)$ is the only position-dependent quantity in $T^0_0$, its perturbation may be written
\begin{equation}
  -\delta \rho = \delta T^0_0 = T^0_0 - {\bar T}^0_0
  = -\frac{m^4}{a^4} \int d^3\vec u \sqrt{a^2+|\vec u|^2} {\bar F}(|\vec u|)
  \left[\frac{F(x,U)}{{\bar F}(|\vec u|)} - 1 \right],
\end{equation}
where the background, homogeneous distribution function $\bar F$ is assumed to be a function of the magnitude of $\vec u$ alone.  The quantity in square brackets is the only factor that depends on the angular components $\hat u = \vec u / |\vec u|$, and angular integration projects out its monopole, i.e., 
\begin{equation}
  \delta \rho
  = \frac{m^4}{a^4} 
     \!\!\int d|\vec u| 4\pi |\vec u|^2 \sqrt{a^2+|\vec u|^2} 
         {\bar F}(|\vec u|) \delmono(x,|\vec u|),
           \label{e:drho_int_u}
\end{equation}
where $\delmono := \int d^2 \hat u/(4\pi) \left[F(x,U)/{\bar F}(|\vec u|)  -  1 \right]$ is the monopole.  Since all non-interacting particles beginning at the same $x$ and $U$ will move in the same way, regardless of their masses, all decoupled HDM species have the same $\delta(x,\vec u) = F(x,U) / {\bar F}(|\vec u|)-1$.  This allows us to reconstruct the perturbations of the individual component species by simply reweighing the flow perturbations.  That is, 
\begin{equation}
  \delta\rho^{(s)}
  =
  \frac{g^{(s)} (m^{(s)})^4}{a^4}
  \int d|\vec u| 4\pi |\vec u|^2 \sqrt{a^2+|\vec u|^2}
         {\bar F}^{(s)}(|\vec u|) \delmono(x,|\vec u|)
         \label{eq:deltarhos}
\end{equation}
for the density perturbation of the species~$s$.

In summary, we have generalized the method of effective distributions of Ref.~\cite{Bayer:2020tko} to the case of multiple HDM species with arbitrary masses, temperatures, and distribution functions, at all times after the species have decoupled.  We have shown its applicability to relativistic as well as non-relativistic HDM, and demonstrated how to recover the density perturbations of the component HDM species.  A key task of perturbation theory for HDM, then, is to determine $\delta(x,\vec u)$.  References~\cite{Chen:2020bdf,Chen:2022cgw}, summarized in Sec.~\ref{subsec:bkg:multi-fluid_perturbation_theories}, did this for a discrete set of $|\vec u|$.  We next consider how to choose these velocities.

\subsection{Discrete momenta and Gauss-Laguerre quadrature}
\label{subsec:eff:discrete_p_and_GLQ}

Section~\ref{subsec:bkg:gauss-laguerre_quadrature} summarizes the GLQ method for approximating the integral of an exponentially-decaying function in some parameter $x$, a set that includes  the Bose-Einstein, Fermi-Dirac, and Maxwell-Boltzmann distribution functions.  The complication is that each $F_{\rm \HDM}^{(s)}$ has its own exponential decay behaviour, $\exp(-\mhdms |\vec u| / \Thdms)$.  We therefore need to choose an effective mass $\mehdm$ and temperature constant $\Tehdm$ for our effective HDM particle, such that for  $q := \mehdm |\vec u| / \Tehdm$, $\Fehdm(q)$  declines as $\exp(-q)$ in in the range of $q$ contributing the most to the density, so that we can effectively apply the GLQ method.
 
 There is no general prescription for selecting $\mehdm$ and $\Tehdm$.  Motivated by the fact that  both the mean energy density and number density scale as $(1+z)^3$ at late times $z\sim 0$ to an excellent approximation, we define for $\Nhdm$ species the effective HDM mass to be
\begin{equation}
  \mehdm
  =
  \frac{\sum_{s=0}^{\Nhdm-1} \bar\rho_{\rm \HDM,0}^{(s)}
  }{\sum_{s=0}^{\Nhdm-1} \bar n_{\rm \HDM,0}^{(s)}},
  \label{e:def_mehdm}
\end{equation}
to ensure that the contribution of each $s$ to $\mehdm$ is weighted by its contribution to the total density. 
We further set the effective temperature according to
\begin{equation}
  \Nhdm \mehdm \Tehdm = \sum_s \mhdms \Thdms,
  \label{e:def_Tehdm}
\end{equation}
i.e., $\Tehdm$ is the mass-weighted average temperature.  Note that the effective distribution function will not, in general, be any equilibrium distribution function, so $\Tehdm$ does not imply any physical system in thermal equilibrium at that temperature.  

Reference~\cite{Bayer:2020tko} instead set $\mehdm$ to the largest of the individual-species masses.  While this is a reasonable choice for their case of interest, in which all $\Thdms$ and $\Tehdm$ are equal, it is no longer optimal if the heaviest species is also several times colder than the rest, causing its contribution to the total density to be subdominant.  Another choice is to demand that $\Fehdm(q)$ be proportional to $\exp(-q)$ at large $q$, which immediately sets $\mehdm/\Tehdm$ equal to the smallest value of $\mhdms/\Thdms$ amongst all species $s$.  This choice is however also not optimal in general, as the lightest and hottest species generally contribute the least to small-scale clustering.  Moreover, reducing the mass of this species will compress the distribution functions of the more dominant species into a smaller range of $q$, leading to larger errors.

We emphasize that the effective distribution function technique is most effective when all HDM species have similar mass-to-temperature ratios.  In this regime, all $\mehdm/\Tehdm$ choices above are likely to give similar results:  we have tested this proposition by increasing $\mehdm/\Tehdm$ by a factor of two relative to \EQS{e:def_mehdm}{e:def_Tehdm}, for a three-neutrino model with masses $42$~meV, $43$~meV, and $65$~meV, and found a similar accuracy.  Outside of this regime, the effective distribution function technique remains mathematically valid, but accuracy will require a large number of quadrature points.  Henceforth we fix $\mehdm$ and $\Tehdm$ as per \EQS{e:def_mehdm}{e:def_Tehdm}.

Given these definitions, we can now change the variables in $\Fehdm$ from $|\vec u|$ to $q = \mehdm |\vec u| / \Tehdm$, i.e., 
\begin{equation}
  \Fehdm(q)
  =
  \frac{1}{\mehdm^4}
  \sum_{s=0}^{\Nhdm-1} \ghdms (\mhdms)^4
  \Fhdms\!\!\left(q \frac{\Tehdm}{\Thdms} \frac{\mhdms}{\mehdm} \right).
  \label{e:Fehdm}
\end{equation}
Section~\ref{subsec:bkg:gauss-laguerre_quadrature} then tells us that an integrand should be evaluated at values $q_\alpha$, for $0 \leq \alpha \leq \Nglq-1$  equal to the $\Nglq$ roots of the $\Nglq$th Laguerre polynomial, for a positive integer $\Nglq$.  Taking the Fourier transform of $\vec x$ and suppressing the time-dependence of $\delmono$, we may now approximate the perturbed density of \EQ{e:drho_int_u} as
\begin{equation}
\begin{aligned}
  \delta\rho^{\vec k}
  & =
  \frac{4\pi \mehdm \Tehdm^3}{a^3}
  \int dq\,q^2\sqrt{1 + \frac{\Tehdm^2 q^2}{\mehdm^2 a^2}}
  \Fehdm(q) \delmono^{\vec k}(q)
  \label{e:drho_k}
  \\
 & \approx
  \frac{4\pi \mehdm \Tehdm^3}{a^3}
  \sum_{\alpha=0}^{\Nglq-1} w_\alpha e^{q_\alpha} q_\alpha^2
  \sqrt{1 + \frac{\Tehdm^2 q_\alpha^2}{\mehdm^2 a^2}}
  \Fehdm(q_\alpha) \delta^{\vec k}_{\alpha 0}
  \\
&  =
  \sum_\alpha \bar\rho_\alpha \delta^{\vec k}_{\alpha 0}
  ~\textrm{where}~
  \bar\rho_\alpha := \tfrac{4\pi \mehdm \Tehdm^3}{a^3}
  \sum_\alpha w_\alpha e^{q_\alpha} q_\alpha^2
  \sqrt{1 + \tfrac{\Tehdm^2 q_\alpha^2}{\mehdm^2 a^2}}
  \Fehdm(q_\alpha),
\end{aligned}
\end{equation}
and $\delta_{\alpha 0}$ is evaluated at $q_\alpha$, that is, $|\vec u| = q_\alpha \Tehdm/\mehdm$, corresponding to momentum $\tau_\alpha = q_\alpha \Tehdm$.  The weight $w_\alpha$ is given by \EQ{e:def_w_GLQ}.  Recall also our convention of Sec.~\ref{subsec:bkg:multi-fluid_perturbation_theories} that a wave number superscript denotes a functional dependence upon that wave number.

These GLQ density perturbations $\delta_\alpha^{\vec k}$ correspond precisely to the flow perturbations of Refs.~\cite{Chen:2020bdf,Chen:2022cgw}, summarized in  Sec.~\ref{subsec:bkg:multi-fluid_perturbation_theories}.  The effective HDM method has shown us how to use the same flow perturbations for an arbitrary collection of HDM particles, while GLQ has shown us how to choose the flow momenta $\tau_\alpha$ efficiently.  We may use these same $\delta_\alpha^{\vec k}$ to recover the density perturbations of individual species via
\begin{equation}
  \delta\rho^{(s)}
  \approx
  \tfrac{4\pi \ghdms\Tehdm^3(\mhdms)^4 }{\mehdm^3 a^3} 
  \!\sum_{\alpha=0}^{\Nglq-1} \!\!\! w_\alpha q_\alpha^2 e^{q_\alpha}
  \!\sqrt{1 + \tfrac{\Tehdm^2 q_\alpha^2}{\mehdm^2 a^2}}
  \Fhdms\!\!\left(q_\alpha \tfrac{\Tehdm}{\Thdms} \tfrac{\mhdms}{\mehdm} \right)
  \delta_\alpha^{\vec k},
  \label{e:drho_species_glq}
\end{equation}
which is the discretized version of \EQ{eq:deltarhos} within the GLQ scheme.

\subsection{Clustering and free-streaming regimes}
\label{subsec:eff:clustering_and_free-streaming_regimes}

Our main goal is to quantify HDM clustering through its monopole density perturbation $\delta_{\alpha,\ell=0}^k$.  We approach this goal by considering $\delta_{\alpha,0}^k$ in two different regimes, the clustering and free-streaming regimes.  In the clustering regime, at length scales much larger than the free-streaming scale, each HDM flow clusters like CDM, so $\delta_{\alpha,0}^k = \delta_{\rm m}^k$.  In the free-streaming regime, at small scales, HDM clustering is strongly suppressed with respect to the total matter clustering.  For linear HDM, $\delta_{\alpha,0}^k = (\kfsa/k)^2 \delta_{\rm m}$, where the free-streaming wave number of \EQ{e:kfs_alpha} separating the clustering and free-streaming regimes can be written
\begin{equation}
  \kfsa(a)^2
  = \frac{3 \Om(a) \Hc^2}{2 v_\alpha^2}
  = \frac{3 \Omo \Hco^2}{2 a v_\alpha^2}
  = \frac{3 \Omo \Hco^2 \mehdm^2 a}{2 \tau_\alpha^2}
  = \frac{3 \Omo \Hco^2 \mehdm^2 a}{2 \Tehdm^2 q_\alpha^2},
  \label{e:kfsa}
\end{equation}
as shown in Refs.~\cite{Ringwald:2004np,Wong:2008ws,Chen:2020kxi,Chen:2020bdf}.  They also confirm the $\sim 10\%$ accuracy of the interpolation $\delta^k_{\alpha,0} \approx (1 + k/\kfsa)^{-2} \delta_{\rm m}^k$ between the two regimes.

We are especially interested in late-time non-relativistic clustering,  $q \Tehdm \ll a \mehdm$, for which the momentum integral in \EQ{e:drho_k} simplifies to $\int dq\,q^2 \Fehdm(q) \delta^{\vec k}(q)$.  The clustering limit is simple, as $\delta^{\vec k}(q) \approx \delta^k_{\rm m}$ is momentum-independent and can be factored out of the integral.  Thus $\delta\rho_{\rm \HDM, clus}^k \approx \bar\rho_{\rm \HDM}\delta^k_{\rm m}$.  The momentum integral is then the same one that we need to compute the average HDM density,
\begin{equation}
  \bar\rho_{\rm \HDM}(a)
  =
  \frac{4\pi \mehdm \Tehdm^3}{a^3}
  \int dq \, q^2 \Fehdm(q)
  \approx
  \frac{4\pi \mehdm \Tehdm^3}{a^3}
  \sum_{\alpha=0}^{\Nglq-1} w_\alpha q_\alpha^2 \Fehdm(q_\alpha).
\end{equation}
Thus, in the clustering limit, we must choose $\Nglq$ high enough for $\int dq \, q^2 \Fehdm(q) \approx   \sum_{\alpha=0}^{\Nglq-1} w_\alpha q_\alpha^2 \Fehdm(q_\alpha)$ to our desired level of accuracy.

The free-streaming regime is more complicated.  As a rough estimate using \EQ{e:kfsa}, we may substitute $\delta^k_{\alpha,0} = (\kfsa^2/k^2)\delta_{\rm m}$ for $\delta_{\rm m}$ in the integral over $q$ in \EQ{e:drho_k}.  Since $\kfsa^2 \propto q_\alpha^{-2}$, the result is proportional to $\int dq \Fehdm(q)$.  This means that, unlike the clustering limit, the free-streaming regime is dominated by lower $q$, and the convergence of GLQ must be considered separately in this limit.  If every single HDM species has a distribution function that is finite down to $q=0$, then the integral $\int dq\,  \Fehdm(q)$ converges.   If however a single bosonic species is present, then $\Fehdm(q) \propto 1/q$ as $q\to 0$, leading to a logarithmic divergence of the integral as the lower integration limit approaches zero.  Thus we cannot speak rigorously of a free-streaming limit in the general case.   Furthermore, the dominance of low-$q_\alpha$ flows means that, at large but finite $k$, the total $\delta^k_{\alpha,0}$ will receive significant contributions from flows that are not yet in the free-streaming regime, that is, $(\kfsa^2/k^2)\delta_{\rm m} \ll \delta^k_{\alpha,0} \lesssim \delta_{\rm m}$.  Thus, to determine the convergence criterion for GLQ, we should instead focus on the intermediate regime between clustering and free-streaming.

To this end, we note that each individual flow has an approximate interpolated solution given by~\cite{Ringwald:2004np}:
\begin{equation}
  \delta^k_{\alpha,0}
  \approx \frac{\delta_{\rm m}^k}{(1+k/\kfsa)^2}
  = \frac{\delta_{\rm m}}{(1 + q_\alpha/q_{\rm cut})^2},
\end{equation}
where at the second equality we have recast the $k/k_{{\rm FS},\alpha}$ in terms of an infrared cutoff defined as $q_{\rm cut}(a,k)^2 := (3\Omo \Hco^2 \mehdm^2 a)/(2 k^2 \Tehdm^2)$.
The expression reproduces the correct behaviors at both $k \ll k_{{\rm FS},\alpha}$ and $k \gg k_{{\rm FS},\alpha}$ limits, making it suitable for the intermediate regime.  Clearly, integrating over $q$ in the monopole density perturbation now returns an expression proportional to $\int dq\, q^2 \Fehdm(q) / (1+q/q_{\rm cut})^2$, which converges at all finite $k$, even for distribution functions $\propto 1/q$ at low $q$, such as the Bose-Einstein distribution. 

Thus we arrive at the convergence criteria for the application of the GLQ scheme.  Suppose we are given an effective HDM whose clustering at $k\leq k_{\rm max}$ and $a \geq a_{\rm min}$ we would like to compute.  We must choose a sufficiently high $\Nglq$ so that the conditions
\begin{equation}
  \int dq \, q^2 \Fehdm(q)
  \approx
  \sum_{\alpha=0}^{\Nglq-1} w_\alpha q_\alpha^2 e^{q_\alpha} \Fehdm(q_\alpha),
    \label{e:glq_error_est_lo_k}
\end{equation} 
and 
\begin{equation}
  \int dq\ \frac{q^2 \Fehdm(q)}{(1+q/q_{\rm cut}(a_{\rm min},k_{\rm max}))^2}
  \approx
  \sum_{\alpha=0}^{\Nglq-1}
  \frac{w_\alpha q_\alpha^2 e^{q_\alpha} \Fehdm(q_\alpha)
  }{(1+q_\alpha/q_{\rm cut}(a_{\rm min},k_{\rm max}))^2}
  \label{e:glq_error_est_hi_k}
\end{equation}
are both satisfied to our desired accuracy.  For practical purposes, we choose $k_{\rm max} / \sqrt{a_{\rm min}} = 10~h/$Mpc.  We may truncate the GLQ series to $\alpha < N_\tau$ flows for $N_\tau < \Nglq$  provided that this truncation keeps the sums on the right hand side within our error threshold.

\subsection{Convergence with \texorpdfstring{$\Nglq$}{Nglq} and \texorpdfstring{$N_\tau$}{Ntau}}
\label{subsec:eff:convergence_with_Nglq_and_Ntau}

\begin{figure}[t]
\begin{center}
  \includegraphics[width=0.99\textwidth]{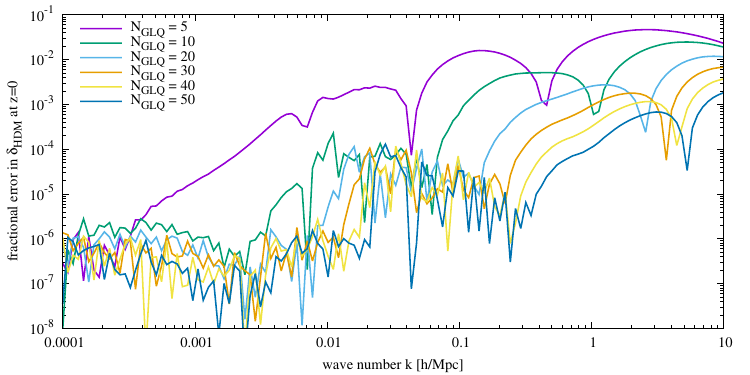}%
  \end{center}
  \caption{
    Convergence of Gauss-Laguerre quadrature for a model with three NO neutrinos
    of total mass $M_\nu=150$~meV and other parameters given by
    \EQ{e:cosmo_params_high-om} for various choices of $\Nglq$.  For each $\Nglq$ considered, we plot the fractional error incurred in the estimate of the effective HDM density monopole $\dhdm$, relative to an estimate of the same using $\Nglq=70$.
    \label{f:d_hdm_convergence_Nglq}  }
\end{figure}

Figure~\ref{f:d_hdm_convergence_Nglq} quantifies the $k$-dependent errors in GLQ as $\Nglq$ is raised, by comparing  estimates of the effective HDM density contrast using various choices of $\Nglq$ relative to a large-$\Nglq (=70)$ estimate.  We have assumed a $\nu\Lambda$CDM model with three NO neutrinos of mass $M_\nu=150$~meV, and other parameters
\begin{equation}
  \Omo h^2 = 0.1518;~\,
  \Obo h^2 = 0.02242;~\,
  A_{\rm s}=2.2\times 10^{-9};~\,
  n_{\rm s} = 0.9665;~\,
  h = 0.6766.
  \label{e:cosmo_params_high-om}
\end{equation}
The CDM+baryon (CB) fluid is evolved using Time-RG perturbation theory, to which the neutrinos respond linearly.  For each choice of $\Nglq$, we use as large an $N_\tau$ as possible while keeping $q_{N_\tau-1}<100$, a convention that we adopt henceforth unless otherwise mentioned.  That is, for our choices of $\Nglq$ of $5$, $10$, $20$, $30$, $40$, $50$, $70$, and $100$, the corresponding settings of $N_\tau$ are, respectively, $5$, $10$, $20$, $29$, $36$, $41$, $50$, and $61$.  As expected, errors grow with $k$ beyond the free-streaming wave number $\kfs = 0.04~h/$Mpc.  Encouragingly, they remain under $0.2\%$ for $\Nglq=50$ and under $1.2\%$ for $\Nglq=20$, meaning that high precision can be attained with a modest number of flows.

We may also use \EQS{e:glq_error_est_lo_k}{e:glq_error_est_hi_k} to estimate the error in GLQ for given $\Nglq$ by comparing a slow but accurate numerical quadrature of the left-hand side to GLQ on the right-hand side.  Consider the high-$k$ error in particular, with $q_{\rm cut}$ specified by $k_{\rm max}=10~h/$Mpc and $a_{\rm min}=1$.  For $\Nglq$ of $5$, $10$, $20$, $30$, $40$, and $50$, \EQ{e:glq_error_est_hi_k} estimates errors of $0.84\%$, $0.82\%$, $0.74\%$, $0.67\%$, $0.59\%$, and $0.52\%$, respectively, compared with actual errors of $2.3\%$, $1.9\%$, $1.1\%$, $0.67\%$, $0.37\%$, and $0.18\%$ in Fig.~\ref{f:d_hdm_convergence_Nglq}.   Thus the error estimate of \EQ{e:glq_error_est_hi_k} is accurate at the order-of-magnitude level, providing a rough guide to the necessary $\Nglq$ for a given error tolerance. 
Meanwhile, at low $k$, the error estimates of \EQ{e:glq_error_est_lo_k} for $\Nglq$ of $5$ and $10$, respectively $5\times 10^{-4}$ and $4\times 10^{-6}$, somewhat overestimate the errors of Fig.~\ref{f:d_hdm_convergence_Nglq} at $k \sim 10^{-3}~h/$Mpc.  However, as $\Nglq$ is increased in the figure, the error seems to hit a floor around $10^{-7}$.  A larger floor $\sim 10^{-5}$ is also evident at intermediate scales $k \sim 0.1~h/$Mpc.  As these are well within our error budget and subdominant to high-$k$ errors, we do not investigate them further.

\begin{figure}[t]
\begin{center}
\includegraphics[width=0.99\textwidth]{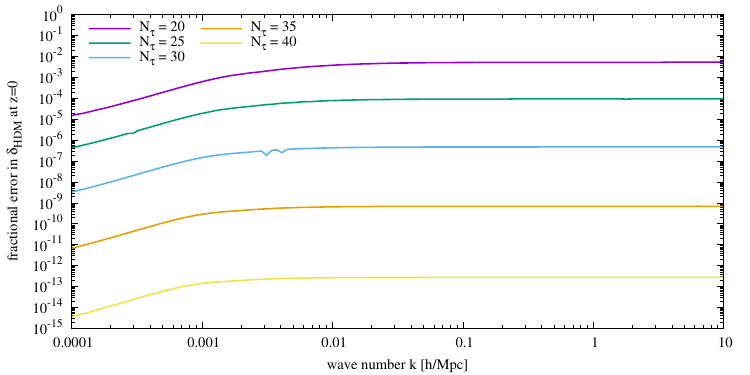}
\caption{Convergence of GLQ with $N_\tau$, for a model with three neutrinos of total mass $M_\nu=150$~meV in the normal mass hierarchy, and other parameters given by \EQ{e:cosmo_params_high-om}.  $\Nglq=100$ is fixed, while the number of flows $N_\tau$ is varied, and the result is compared with $N_\tau=60$.}
\label{f:d_hdm_convergence_Ntau}
\end{center}
\end{figure}

Figure~\ref{f:d_hdm_convergence_Ntau} considers the convergence of GLQ as we vary $N_\tau$ in the estimate of the effective HDM monopole density contrast.
With $\Nglq=100$ fixed, $N_\tau=60$ suffices to reach $q_{N_\tau-1}=95$, and we test smaller values of $N_\tau$ against the case of $N_\tau=60$.  Evidently, for every choice of $N_\tau$ shown, the corresponding estimate of $\delta_{\rm HDM}$ has converged to better than $1\%$.  Even a modest increase in $N_\tau$ rapidly decreases the error across the whole $k$-range;  At $N_\tau \geq 45$, the errors fall below the numerical precision of $\sim 10^{-16}$ and are thus not shown in the figure.
The corresponding low-$k$ error estimates from comparing the two sides of \EQ{e:glq_error_est_lo_k} are $5\times 10^{-3}$ for $N_\tau=20$, $9\times 10^{-5}$ for $N_\tau=25$, $5\times 10^{-7}$ for $N_\tau=30$, $7\times 10^{-10}$ for $N_\tau=35$, and $2\times 10^{-13}$ for $N_\tau=40$, about an order-of-magnitude larger than the the  low-$k$ errors in Fig.~\ref{f:d_hdm_convergence_Ntau}. 
 At high $k$, the error estimates of \EQ{e:glq_error_est_hi_k} are: $9\times 10^{-5}$ for $\Nglq=20$; $7\times 10^{-7}$ for $\Nglq=25$; $2\times 10^{-9}$ for $\Nglq=30$; $1.5 \times 10^{-12}$ for $\Nglq=35$; and $3\times 10^{-16}$ for $\Nglq=40$.  These are about two orders of magnitude smaller than the high-$k$ errors of Fig.~\ref{f:d_hdm_convergence_Ntau}, but scale similarly with $\Ntau$, demonstrating that \EQS{e:glq_error_est_lo_k}{e:glq_error_est_hi_k} are reasonable approximate guides to GLQ convergence with $\Ntau$.

\subsection{Non-linear perturbation theory: \FFTMII{}}
\label{subsec:eff:non-linear_perturbation_theory_fftmii}

In principle, the implementation of GLQ and EHDM in \FFTM{} is straightforward.  EHDM simply means replacing the Fermi-Dirac distribution for neutrinos by the effective distribution function $\Fehdm(q)$ of \EQ{e:Fehdm}.  GLQ requires $\tau_\alpha = q_\alpha \Tehdm$ for each flow $\alpha$, and its corresponding late-time density fraction is proportional to $w_\alpha q_\alpha^2 \exp(q_\alpha) \Fehdm(q_\alpha)$, as discussed in
Sec.~\ref{subsec:eff:discrete_p_and_GLQ}.
However, problems arise for the smallest $q_\alpha$.

Reference~\cite{Chen:2022cgw} encountered  numerical instabilities in the \FFTM{} perturbation theory applied to massive neutrinos in the high-$k$ and high-$\ell$ regime.  A stability threshold~$\kThr$ had to be introduced, above which evolution of the perturbations was no longer tracked, and $\kThr$ was reduced  dynamically as instabilities caused a reduction of the integration step size to below $\Delta\eta=10^{-6}$.  For the fiducial model used, with $\Ono h^2 = 0.005$, Ref.~\cite{Chen:2022cgw} found it possible to stabilize \FFTM{} up to $\kThr \gtrsim 3~h/$Mpc by truncating the Legendre moment expansion of the power spectrum input to the mode-coupling integrals, $\ell < \Nmunl$, meaning that $\Aa{acd,bef,\ell}{k}$ was nonzero only for $\ell < 2\Nmunl-1$.  A choice of $\Nmunl$ as high as $8$ was found to be computationally tractable, but $\Nmunl=6$ achieved a reasonable balance between precision and computational cost.  Therefore, following~\cite{Chen:2022cgw}, we adopt the choice $\Nmunl=6$ henceforth, unless stated otherwise.

However, applying \FFTM{} with $\Nmunl=6$ to a much broader range $\Ono h^2 \leq 0.01$, Ref.~\cite{Upadhye:2023bgx} found that this truncation alone was inadequate for the smallest~$\tau$ and largest neutrino masses, that is, for the smallest velocities.  Numerical instabilities affected nearly $20\%$ of their sample of $101$ runs.  They were able to stabilize \FFTM{} for this larger mass range by introducing a further truncation of the $\ell$ range of the bispectrum integrals $\Ia{acd,bef,\ell}{k}$ and the mode-coupling integrals $\Aa{acd,bef,\ell}{k}$, restricting $\ell < \Nmuai$.  They defined stability as the integration being able to reach $z=0$ with $\kThr \geq 1.2~h/$Mpc, a definition which we adopt here.   The choice of $\Nmuai$ = $4$ or $5$ was found to stabilize the perturbation theory, at the cost of an additional error of $1\%-2\%$, which they quantified by comparison to higher $\Nmuai$.  While the error associated with this truncation was larger for lower $\Ono h^2$, truncation was only necessary for $\Ono h^2 > 0.006$.

GLQ exacerbates this instability by requiring more low-velocity bins, necessary for accurately predicting the small-scale clustering of HDM.  We find that even for $\Mnu=59$~meV, the minimum allowed value in the normal ordering, \FFTM{} is unstable.  Furthermore, reducing $\Nmuai$ to accommodate the lowest-$\alpha$ flows will introduce unacceptable errors into the larger $\alpha$; that is, the $\Nmuai$ truncation is too aggressive for our purposes.  Thus we allow each flow $\alpha$ to have its own truncation, $\ell < \Nmuai^{(\alpha)}$, for its $\Ia{acd,bef,\ell}{k}$ and $\Aa{acd,bef,\ell}{k}$.  A full exploration of the parameter space of all $\Nmuai^{(\alpha)}$ would be prohibitively expensive computationally, and we do not consider it here.  However, for a modest GLQ order $\Nglq=20$, we find that reducing only $\Nmuai^{(0)}$ is sufficient to ensure the stability of models with densities $\Ono h^2 \lesssim 0.003$, while larger $\Nglq$ can be reached for smaller neutrino masses.  Henceforth, unless otherwise mentioned, we only discuss non-linear perturbative results for which stability may be achieved by reducing  $\Nmuai^{(0)}$ alone, leaving all others at their maximum value of $2\Nmunl-1 = 11$.

\begin{figure}[t]
\begin{center}
  \includegraphics[width=0.49\textwidth]{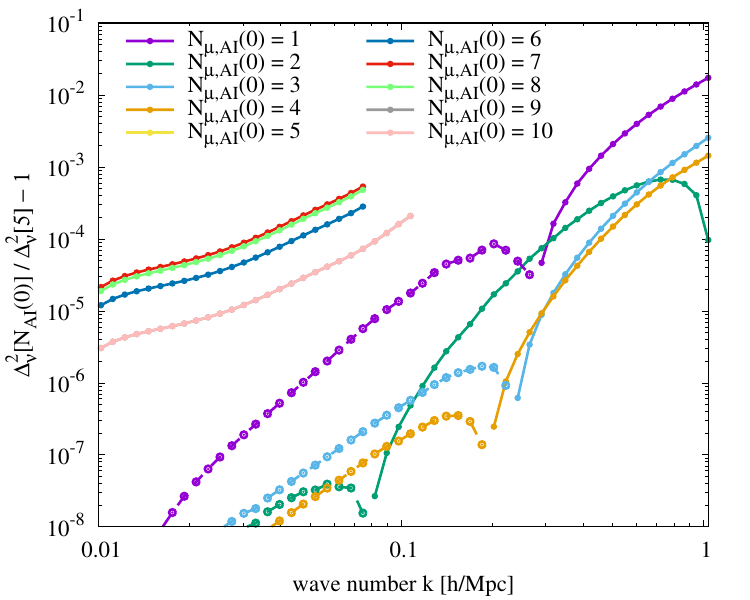}
  \includegraphics[width=0.49\textwidth]{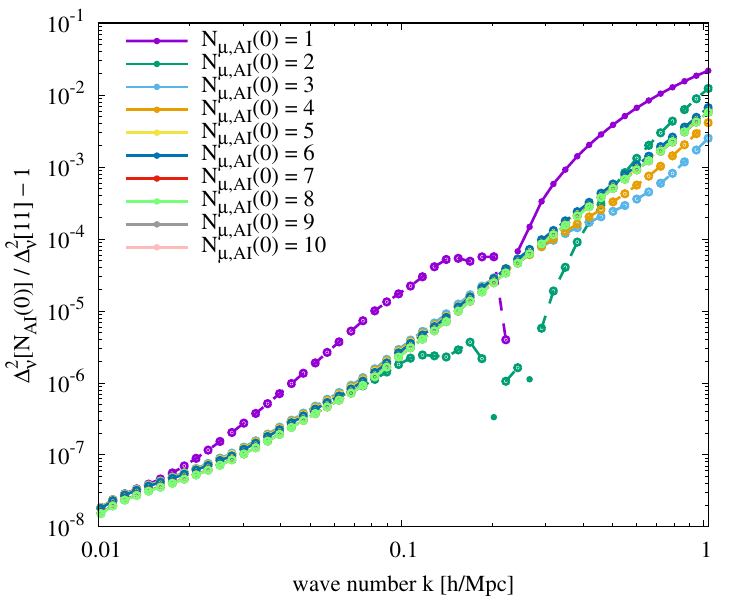}%
  \caption{
    Fractional differences in the $z=0$ neutrino power spectra computed using various choices of $\Nmuai^{(0)}$ relative to that computed using the largest stable $\Nmuai^{(0)}$ value,  
 for a massive neutrino model with $\Mnu=59$~meV.
    {\it Left}: Normal mass ordering.  The largest $\Nmuai^{(0)}$ achieving
    stability is 5. 
    {\it Right}: Degenerate mass ordering.  All $\Nmuai^{(0)}$ are stable, so we
    have used $\Nmuai^{(0)}=11$ as the reference for comparison.
    \label{f:Nmuai_errors}
  }
  \end{center}
\end{figure}

Figure~\ref{f:Nmuai_errors} compares the power spectra for the $\Mnu=59$~meV normal and degenerate mass orderings as $\Nmuai^{(0)}$ is varied.  Since our $\Nmuai^{(0)}$ truncation is less aggressive than the global $\Nmuai$ truncation of Ref.~\cite{Upadhye:2023bgx}, we find considerably smaller errors.  Even $\Nmuai^{(0)}=1$ reaches $2.2\%$ accuracy all the way to $k=1~h/$Mpc for the degenerate ordering, with smaller errors for the normal ordering.  Furthermore, since these errors decrease with increasing neutrino mass, and Fig.~\ref{f:Nmuai_errors} considers the smallest allowed $\Mnu$, we may use $2.2\%$ as an upper bound for all larger masses in the case of $\Nmuai^{(0)}=1$; $1.2\%$ in the case of $\Nmuai^{(0)}=2$; and $0.25\%$ in the case of $\Nmuai^{(0)}=3$, with higher $\Nmuai^{(0)}$ consistently under $1\%$ for both mass orderings.  Thus we regard this $\Nmuai^{(0)}$ truncation as sufficiently accurate for our purposes, though its extension to higher masses than those considered in this article will require further consideration.  Henceforth, we refer to this new version of \FFTM{}, implementing EHDM and GLQ, with  both $\Nmunl$ and $\Nmuai^{(\alpha)}$ truncations, as \FFTMII{}.

\subsection{Numerical accuracy}
\label{subsec:eff:numerical_accuracy}

We conclude this section by assessing the numerical accuracy of GLQ and EHDM in  perturbation theory.  We consider $\nu\Lambda$CDM models with the cosmological parameters of \EQ{e:cosmo_params_high-om} and $\Mnu=150$~meV neutrinos in either the normal or degenerate mass ordering. In the case of linear perturbation theory, we compare our results to the \classcode{} code, in which we set the three neutrino masses individually.  Our \classcode{} runs fix all non-cold dark matter (NCDM) tolerances to $10^{-9}$ and set {\tt l\_max\_ncdm=500}.

\begin{figure}[t]
\begin{center}
  \includegraphics[width=0.99\textwidth]{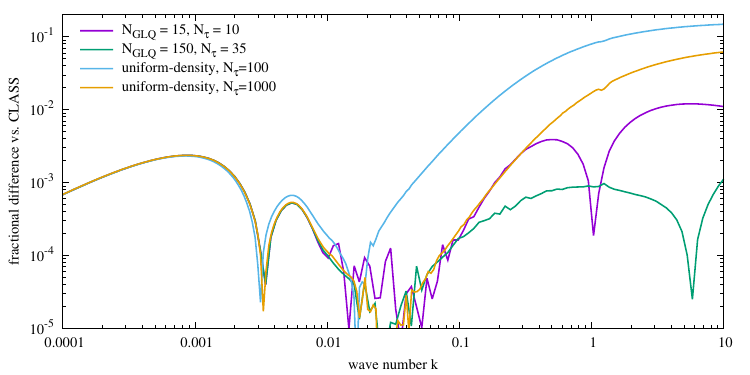}%

  \caption{
    Comparison of HDM densities $\dhdm$, computed using the linear perturbation
    theory of Sec.~\ref{subsec:bkg:multi-fluid_perturbation_theories} with GLQ
    or uniform-density binning, to \classcode{}.
    We consider three degenerate-mass neutrinos with $M_\nu = 150$~meV at $z=0$.
    \label{f:acc_vs_class}
  }
  \end{center}
\end{figure}

Figure~\ref{f:acc_vs_class} considers the DO case in order to facilitate comparison with the degenerate-mass multi-fluid perturbation theory of Ref.~\cite{Chen:2020bdf}, which uses uniform-number-density neutrino bins.  It demonstrates that GLQ is far more efficient than uniform bins.  The number of flows, $N_\tau$, is directly proportional to the computational cost, since the neutrinos are the most computationally-demanding part of the linear perturbative calculation.  Evidently, GLQ with $N_\tau=35$ has a $k\geq 1~h/$Mpc error two orders of magnitude below that of uniform binning with $N_\tau=100$.  GLQ even outperforms the thousand-bin calculation by more than an order of magnitude.  Note that this $N_\tau=35$ GLQ curve is most closely comparable to the $\Nglq=40$ curve in Fig.~\ref{f:d_hdm_convergence_Nglq}, which used $N_\tau=36$.  Both have errors $\lesssim 0.1\%$.

\begin{figure}[t]
\begin{center}
  \includegraphics[width=0.99\textwidth]{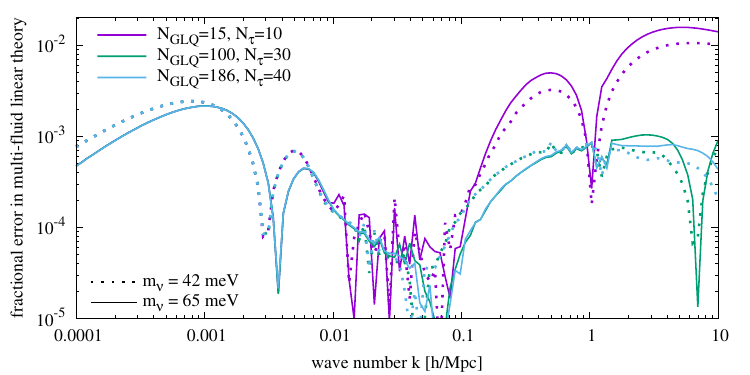}%
  
  \caption{
    Comparison of HDM densities $\dhdm$ computed using the linear perturbation
    theory of Sec.~\ref{subsec:bkg:multi-fluid_perturbation_theories}
    implementing GLQ and EHDM, with the output of 
    \classcode{} at $z=0$.  The model has three NO neutrinos with $\Mnu=150$~meV.  The
    $42$~meV ($65$~meV) neutrino is shown using dotted (solid) lines; 
    errors in the $42$~meV and $43$~meV neutrinos are nearly identical.
    \label{f:acc_FLPT_Mnu150_NO_vs_class}
  }
  \end{center}
\end{figure}

Next, we consider individual neutrino density monopoles in the NO case.  Figure~\ref{f:acc_FLPT_Mnu150_NO_vs_class} demonstrates the accuracy of the combined GLQ and EHDM methods compared with \classcode{}, in which the three neutrinos are considered separately.  Even a modest number of GLQ points, $\Nglq=15$, agrees with \classcode{} to $<2\%$ across the entire $k$ range for the both the light and the heavy neutrinos separately, with somewhat larger errors for the more massive neutrino.  Larger $\Nglq$ reduces these errors to $\leq 0.1\%$ for $k \geq 0.002~h/$Mpc.  Increasing $\Nglq$ beyond $\sim 100$ yields no discernible improvements to the accuracy, showing that the EHDM momentum resolution is no longer a dominant source of error.  We thus confirm that the EHDM method is fundamentally sound and that GLQ is accurate for $k \leq 10~h/$Mpc.

 \begin{figure}[t]
 \begin{center}
  \includegraphics[width=0.99\textwidth]{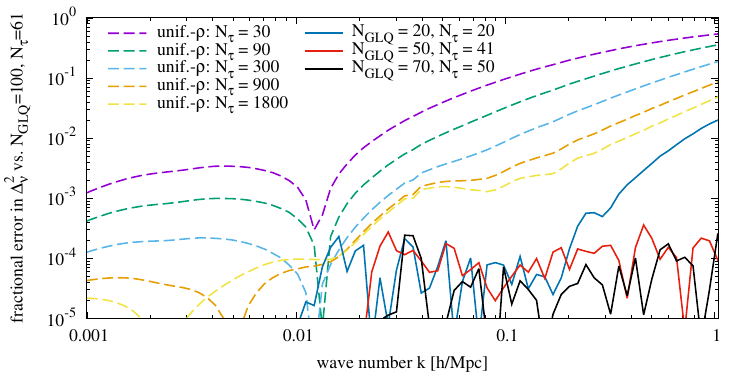}%
  \caption{
    Fractional errors in $\DDnu(k)$ at $z=0$ computed using uniform-density momentum bins (dashed)
    and GLQ bins (solid) relative to a reference model with $\Nglq=100$ and $\Ntau=60$.  The neutrino model contains $\Mnu=150$~meV NO masses.  
    \label{f:FFTMII_vs_FFTM}
  }
  \end{center}
\end{figure}

 Finally, we compare uniform-density three-species binning to GLQ and EHDM in non-linear perturbation theory, for three NO neutrinos with $\Mnu=150$~meV.  The computational expense of \FFTM{} rises in proportion to $\Nmunl^6$, so we restrict ourselves in this subsection to $\Nmunl=3$, the minimum value recommended by Ref.~\cite{Chen:2022cgw}.  In the calculation with uniform-density bins, each of the three neutrinos is tracked using $\Ntau/3$ bins of equal density.  Thus bins corresponding to different neutrino species have different densities.   Since we showed GLQ to have converged by $\Nglq=100$, we use the GLQ power spectrum with $\Nglq=100$ and $\Ntau=61$ as the reference model against which all others are compared. 

 Figure~\ref{f:FFTMII_vs_FFTM} compares several neutrino power spectrum computations using uniform-density as well as GLQ binning methods.  Up to a numerical noise at the $\sim 0.01\%$ level, $\Nglq$ of $50$ and $70$ are nearly identical to the reference model for $k \leq 1~h/$Mpc.  Even $\Nglq=20$ is consistent with that noise up to $k=0.2~h/$Mpc and has $\leq 2\%$ errors up to $k=1~h/$Mpc.  At large scales, $k \leq 0.3~h/$Mpc, the highest-resolution uniform-density binning agrees with all of the GLQ calculations at the percent level.  However, uniform-density binning converges slowly at high $k$, where a high resolution of the smallest momenta is essential to an accurate computation.  At $k=1~h/$Mpc, the lowest-resolution GLQ, with $\Nglq=\Ntau=20$, is more accurate than even the highest-resolution uniform-density binning using nearly a hundred times as many bins, illustrating the advantages of Gauss-Laguerre quadrature.  We also see that the two highest-resolution uniform-density calculations disagree by $\geq 1\%$ for $k \geq 0.4~h/$Mpc and by $\approx 4\%$ at $k=1~h/$Mpc, justifying our use of GLQ for the reference model.  We therefore conclude that \FFTMII{} with EHDM and GLQ has converged at about the percent level for $\Nglq$ of $15$ or $20$.  We proceed to apply it to neutrinos and other HDM models.

\section{Results I: Non-linear enhancement of HDM clustering}
\label{sec:re1}

\subsection{Accuracy at low \texorpdfstring{$\Mnu$}{Mnu}: solving a puzzle}
\label{subsec:res:accuracy_at_low_Mnu}

Reference~\cite{Upadhye:2023bgx} encountered a mysterious $\approx 50\%$ small-scale error in \FFTM{} and its companion \cosmicenu{} emulator.  This error at $k \approx 1~h/$Mpc is evident in comparisons with a series of degenerate-mass $\nu\Lambda$CDM simulations conducted by the Euclid code-comparison project in Ref.~\cite{Euclid:2022qde} for $\Mnu$ ranging from $150$~meV to $600$~meV; the remaining parameters are $\Omo h^2 = 0.1432$, $\Obo h^2 = 0.022$, $A_s=2.215\times 10^{-9}$, $n_s= 0.9619$, and $h=0.67$.  This error is considerably larger than the $15\%-20\%$ error expected from the N-body comparison of the original \FFTM{} publication, Ref.~\cite{Chen:2022cgw}.  The simulation of that reference used a small  volume, a box of edge length $128~{\rm Mpc}/h$, in order to reduce shot noise, at the cost of neglecting larger-scale power that could flow down to smaller scales due to non-linear clustering.  This could explain some of the error, but not its $\Mnu$-independence.

Although they were unable to find a conclusive explanation for this $\Mnu$-independent error, Ref.~\cite{Upadhye:2023bgx} suggested three possibilities:
\begin{enumerate}
\item Perturbation theory error.  Non-linear perturbation theory is most accurate on quasi-linear scales, and smaller-scale accuracy would require higher-order perturbative corrections.  However, lighter neutrinos cluster more linearly, so it is difficult to explain how this error remains $\approx 50\%$ while $\Mnu$ is varied by a factor of four.
\item Non-perturbative clustering.  Perturbation theory cannot account for non-perturbative structures such as CDM halos, which are expected to capture some portion of the neutrino population.  However, again, a fourfold change in $\Mnu$ will substantially change the number of neutrinos below the escape velocity of the typical halo, so such capture should be strongly $\Mnu$-dependent.
\item N-body systematic biases.  Reference~\cite{Euclid:2022qde} saw a $30\%-40\%$ small-scale scatter among the different simulation methods.  This scatter could be larger if errors due to imperfect convergence and incorrect initialization, as studied in Ref.~\cite{Sullivan:2023ntz}, are included.
\end{enumerate}

Here, we are able to conclude definitively that this $50\%$ error is actually the combination of two different errors.  \FFTM{} at $k \approx 1~h/$Mpc is indeed breaking down for higher $\Mnu$.  At low $\Mnu$, an inadequate sampling of low neutrino momenta, responsible for most of the small-scale clustering, leads to large errors.  This latter error can be substantially reduced, either by significantly increasing the number of momentum bins, or by switching to a more efficient quadrature method such as GLQ, as we proceed to show.  

\begin{figure}[t]
\begin{center}
  \includegraphics[width=0.49\textwidth]{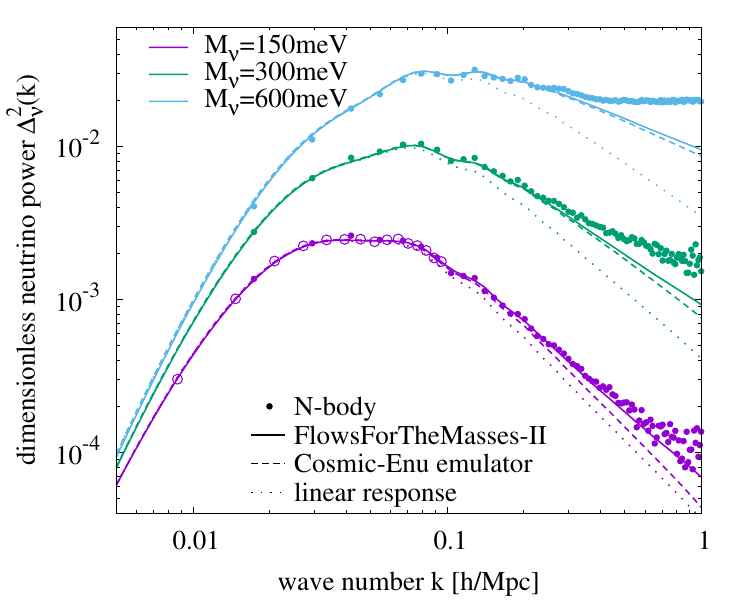}%
  \includegraphics[width=0.49\textwidth]{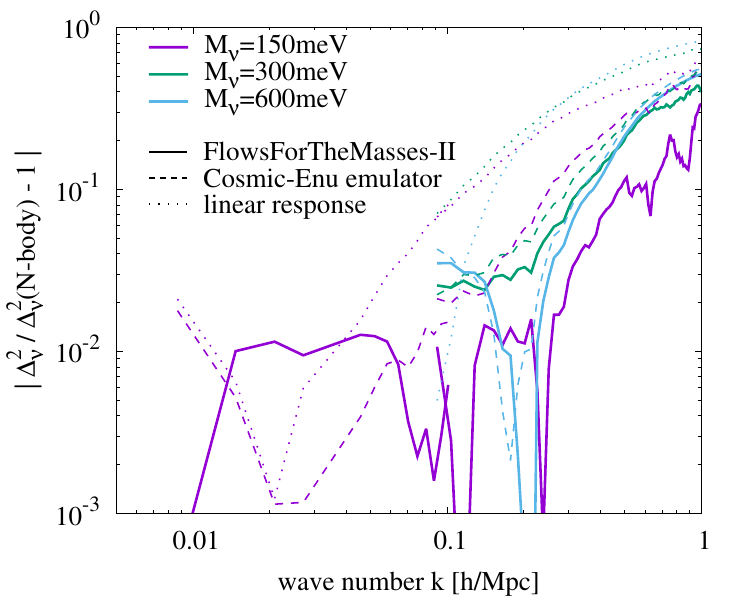}%
  \caption{Predictions of the dimensionless neutrino power spectrum at $z=0$ for the degenerate-ordering $\Mnu=150$~meV, $300$~meV, and $600$~meV $\nu\Lambda$CDM models of
    Ref.~\cite{Euclid:2022qde}, using a variety of computation methods.
{\it Left}: The absolute neutrino power spectrum computed using \FFTMII{} (solid), 
\cosmicenu{} (dashed), linear response
    (dotted), and N-body simulations (points). In the case of N-body predictions, filled points
    correspond to $(512~{\rm Mpc})^3$ simulation volumes and open points
  to a volume of $(1024~{\rm Mpc})^3$ for the $\Mnu=150$~meV model.  {\it Right}: Fractional errors in $\Delta_\nu^2(k)$ relative to the N-body predictions.
    \label{f:Euclid_code_comp_Mnu}
  }
  \end{center}
\end{figure}

Our results for a range of $\Mnu$ are demonstrated in Fig.~\ref{f:Euclid_code_comp_Mnu}.  The \FFTMII{} curves use GLQ with $\Nglq=50$ and $\Ntau=41$, while the \cosmicenu{} curves emulate \FFTM{} using $\Ntau=50$ uniform-density bins.  Up to shot noise in the simulation, this perturbation theory is accurate to $\approx 20\%$ for $\Mnu=150$~meV, $\approx 40\%$ for $\Mnu=300$~meV, and $\approx 50\%$ for $\Mnu=600$~meV.  The switch to GLQ has only a minor impact at the highest mass shown, while the error is dominated by either higher-order or non-perturbative clustering.  By increasing significantly with $\Mnu$, this residual error behaves as expected of a breakdown in perturbation theory.

We find, however, that the numerical instabilities discussed in Sec.~\ref{subsec:eff:non-linear_perturbation_theory_fftmii} become severe for $\Mnu=600$~meV.  That section defined a flow-dependent truncation $\ell < \Nmuai^{(\alpha)}$ of the number of angular modes passed to the non-linear mode-coupling integrals.  Thus far, with $\Mnu \leq 300$~meV, we have found that reducing $\Nmuai^{(0)}$ is sufficient to stabilize the perturbation theory to $k=1.2~h/$Mpc.  However, we find for $\Mnu=600$~meV that we must extend this truncation to the first three flows, reducing  each of $\Nmuai^{(0)}$, $\Nmuai^{(1)}$, and $\Nmuai^{(2)}$ to $3$.  Evidently, GLQ exacerbates the numerical instabilities of \FFTMII{} for large $\Mnu$. 

\begin{figure}[t]
\begin{center}
  \includegraphics[width=0.99\textwidth]{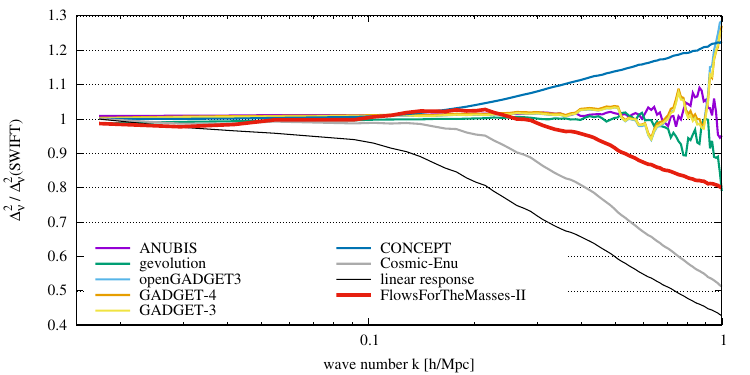}%
  \caption{Comparison of \FFTMII{} in its prediction of the $z=0$ dimensionless neutrino power spectrum $\Delta^2_\nu(k)$  for 
   the degenerate-ordering $\Mnu=150$~meV $\nu\Lambda$CDM model of Ref.~\cite{Euclid:2022qde}, against a
    variety of neutrino simulation methods.  Also shown are the linear
    response power spectrum of Ref.~\cite{Chen:2020bdf} and the \cosmicenu{}
    emulator of Ref.~\cite{Upadhye:2023bgx}.
    \label{f:Euclid_code_comp}
  }
  \end{center}
\end{figure}

Figure~\ref{f:Euclid_code_comp} provides, for the $\Mnu=150$~meV model, further detail about the improved low-momentum sampling in \FFTMII{}.  Here, several different N-body methods~\cite{Teyssier:2001cp,Mauland:2023eax,Adamek:2015eda,Adamek:2016zes,Adamek:2017uiq,Beck:2015qva,Marin-Gilabert:2022ggx,Springel:2005mi,Springel:2008cc,Springel:2020plp,Dakin:2017idt,Dakin:2021ivb} are compared with linear response~\cite{AliHaimoud:2012vj,Chen:2020bdf}, the \cosmicenu{} emulator, and \FFTMII{}; the low-shot-noise {\sc swift} simulation of Ref.~\cite{SWIFT:2023dix}, based upon the $\delta f$ method of Refs.~\cite{Elbers:2020lbn,Elbers:2022tvb}, is used as a reference.  We can now see more clearly that GLQ reduces the \FFTMII{} error to $<10\%$ up to $k=0.5~h/$Mpc and $<20\%$ up to $k=1~h/$Mpc; its errors are now comparable to the scatter among the different N-body methods themselves.  Thus GLQ has reduced the error of \FFTMII{} relative to \FFTM{} and \cosmicenu{} by more than a factor of two.  This represents a significant improvement over the uniform-density-binned codes for $k \gtrsim 0.2~h/$Mpc and over the linear response method for $k \gtrsim 0.05~h/$Mpc.

\begin{figure}[t]
\begin{center}
  \includegraphics[width=0.99\textwidth]{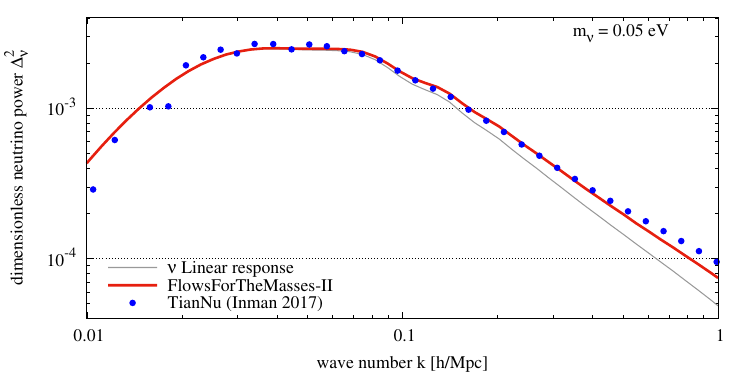}%
  \caption{The $z=0$ dimensionless
    neutrino power spectrum from the TianNu simulation of
    Ref.~\cite{Inman:2016prk}, compared with \FFTMII{} (thick line) and linear response (thin line).  The cosmological model contains one massive neutrino at 50~meV and two massless states.
    \label{f:D2nu_TianNu}
  }
  \end{center}
\end{figure}

The highest-resolution massive neutrino N-body simulation conducted thus far is the TianNu simulation of Refs.~\cite{Yu:2016yfe,Inman:2016prk,Emberson:2016ecv}, which tracked $2.6$ trillion neutrino particles in a cubic box of edge length $1200~{\rm Mpc}/h$.  It approximated the normal mass ordering by simulating a single $50$~meV neutrino, with the remaining two assumed to be massless.  Figure~\ref{f:D2nu_TianNu} compares \FFTMII{}, with $\Nglq=20$ GLQ bins and $\Ntau=20$ momentum bins, to the TianNu power spectrum, finding an accuracy of $21\%$ at $k=1~h/$Mpc.  We have thus confirmed the accuracy of our GLQ-based \FFTMII{} perturbation theory to $\approx 20\%$ for $k \leq 1~h/$Mpc for neutrino masses up to $50$~meV.

We have considered building a new emulator using GLQ momentum bins.  However, the high-$\Mnu$ numerical instabilities noted above for $\Mnu=600$~meV required the adjustment of three separate truncation parameters to resolve, and masses $\Mnu \approx 930$~meV at the upper end of the emulation range may require more.  Individually adjusting this many parameters and demonstrating the insensitivity of the resulting power spectra to their precise values, for every single high-$\Mnu$ model in the training set, would be prohibitively computationally expensive.  Alternatively, we may regard \cosmicenu{} as well-suited to $\Mnu \gtrsim 300$~meV, where the benefits of GLQ are diminishing and where the degenerate mass ordering becomes accurate to $\approx 2\%$, as we shall see in Sec.~\ref{subsec:res:errors_in_DO_approx}.  Then we may use GLQ to construct two separate $\Mnu \leq 300$~meV emulators for the normal and inverted mass orderings.  We leave this for future work.

\subsection{Normal ordering and recovery of individual-species power spectra}
\label{subsec:res:normal_ordering_and_recovery_of_indiv-species_power_spec}

Next, we apply the effective HDM method to the normal hierarchy, with the ultimate goal of verifying the individual-species power spectra implied by \EQ{e:drho_species_glq}.  We compare our results to the {\tt gevolution} N-body simulation of Ref.~\cite{Adamek:2017uiq}, which uses an approximation to the normal mass ordering.  Since multiple neutrino species substantially increase the computational costs of simulations, Ref.~\cite{Adamek:2017uiq} simulated a doubly-degenerate $60$~meV neutrino and a singly-degenerate $80$~meV neutrino, for a total mass $\Mnu=200$~meV.  That is, they neglected the smaller mass splitting, $\Dmmtwo$, and approximated the larger one, $\Dmmthree$, as $0.0028$~eV$^2$.  Their simulation tracked $4096^3$ CDM+baryon particles and $1.7\times 10^{11}$ neutrino particles in a $(2~\rm{Gpc}/h)^3$ box with a force resolution of $0.5~{\rm Mpc}/h$.  

\begin{figure}[t]
\begin{center}
  \includegraphics[width=0.49\textwidth]{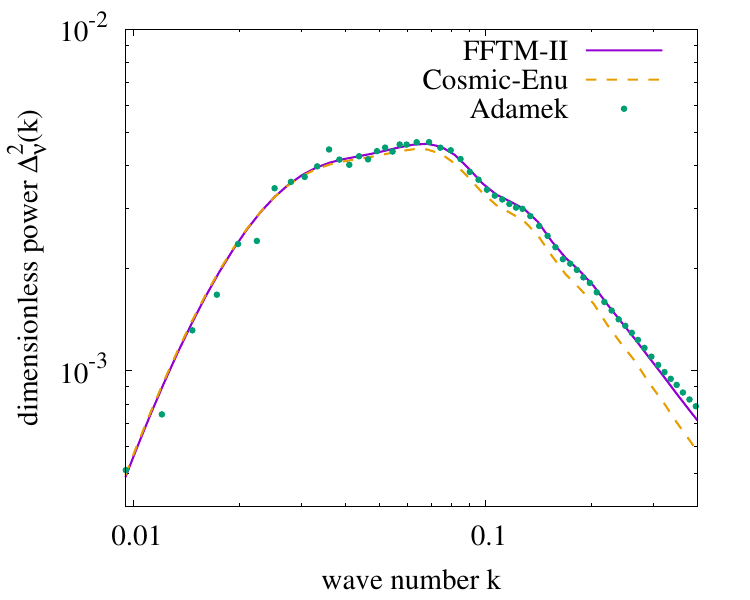}
  \includegraphics[width=0.49\textwidth]{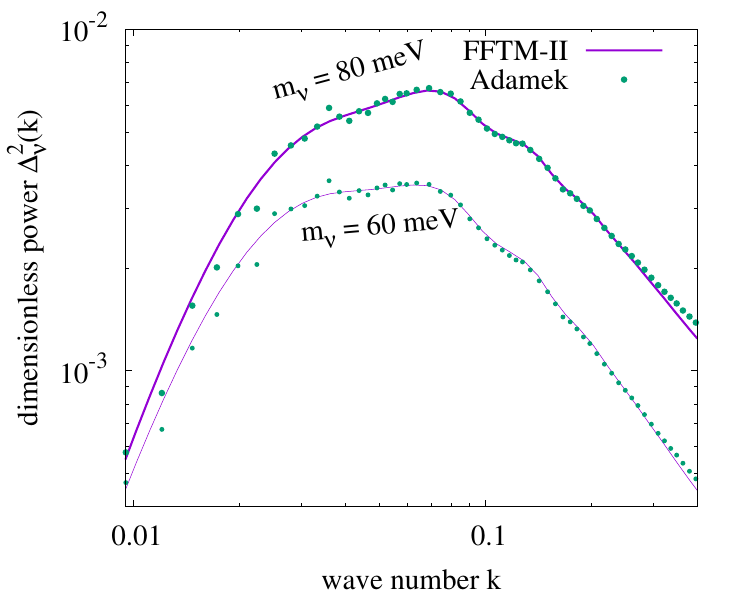}%
  \caption{
    Comparison of the \FFTMII{} neutrino power spectrum at $z=0$, using GLQ momentum binning, to the N-body neutrino simulation results of Ref.~\cite{Adamek:2017uiq}, assuming $\Mnu=200$~meV and an
    approximate normal mass ordering. {\it Left}: Total neutrino power spectrum. For comparison, we plot also the emulated
    \cosmicenu{} power spectrum of Ref.~\cite{Upadhye:2023bgx} for the same $\Mnu$ but assuming a degenerate  ordering.
    {\it Right}: Separate power spectra of the $80$~meV and $60$~meV species.
    \label{f:acc_vs_Nbody}
  }
  \end{center}
\end{figure}

Figure~\ref{f:acc_vs_Nbody} compares \FFTMII{} with $\Nglq=50$ and $\Ntau=41$ to Ref.~\cite{Adamek:2017uiq} up to $k=0.4~h/$Mpc, after which the N-body power spectra become dominated by shot noise.  The other cosmological parameters of this $\nu\Lambda$CDM model are  $\Omo h^2 = 0.142412$, $\Obo h^2 = 0.022032$, $A_s=2.215\times 10^{-9}$, $n_s= 0.9619$, and $h=0.67556$.  Also shown in the left panel is the emulated power spectrum of \cosmicenu{}, which assumes a degenerate neutrino mass ordering.  Relative to \cosmicenu{}, \FFTMII{} represents a nearly fourfold reduction in RMS fractional error  over the range $0.1~h/{\rm Mpc} < k < 0.15~h/{\rm Mpc}$, from $6.2\%$ to $1.6\%$, and over a threefold reduction over $0.35~h/{\rm Mpc} < k < 0.4~h/{\rm Mpc}$, from $24\%$ to $7\%$.

Moreover, \FFTMII{} accurately recovers the power spectra of individual neutrino species, as seen in the right panel of Fig.~\ref{f:acc_vs_Nbody}.  Its RMS fractional errors are comparable to those of the total neutrino power spectrum, though the heavier neutrino species has slightly smaller errors at low $k$ and larger ones at high $k$.  For example, in the $0.1~h/{\rm Mpc} < k < 0.15~h/{\rm Mpc}$ range, the \FFTMII{} error is $0.9\%$ for the $80$~meV species and $2.2\%$ for the $60$~meV species, while in the $0.35~h/{\rm Mpc} < k < 0.4~h/{\rm Mpc}$ range, these errors grow to $8.2\%$ and $5.7\%$, respectively.  This rise in the error of the heavier neutrino suggests a small-scale non-linear effect not captured by perturbation theory.  Also evident from the smallest scales in the same plot is the fact that the $80$~meV neutrino power spectrum is about three times that of the $60$~meV neutrino, consistent with the $\DDnu \propto m_\nu^4$ scaling of Refs.~\cite{Ringwald:2004np,Wong:2008ws}.

\subsection{Errors in the degenerate-ordering approximation}
\label{subsec:res:errors_in_DO_approx}

Bounds on $\Mnu$ commonly make the approximation of a degenerate mass ordering, which reduces the computational cost of their power spectrum computations.  We next investigate the error in the total neutrino power spectrum arising from this approximation.  This error necessarily vanishes in both the clustering limit, when all neutrino masses cluster the same, and the high-$\Mnu$ limit, in which fractional differences between the neutrino masses vanish.  Thus we focus on $\Mnu \leq 300$~meV and $0.01~h/{\rm Mpc} \leq k \leq 1~h/$Mpc.  Since the free-streaming-limit power spectrum for a single neutrino of mass $m_\nu$ scales as $m_\nu^4$ in the linear case~\cite{Ringwald:2004np}, and as an even higher power of $m_\nu$ with non-linear corrections~\cite{Upadhye:2023bgx}, we expect the degenerate mass ordering to underestimate the neutrino power in all cases.  As a numerical example, we consider $\nu\Lambda$CDM cosmologies with the parameters of \EQ{e:cosmo_params_high-om} fixed.

\begin{figure}[t]
\begin{center}
  \includegraphics[width=0.49\textwidth]{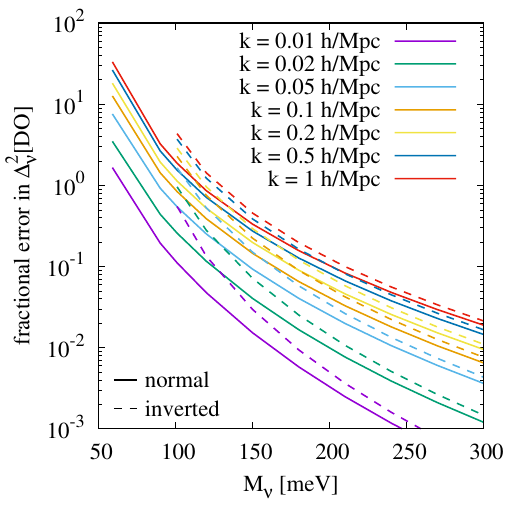}%
  \includegraphics[width=0.49\textwidth]{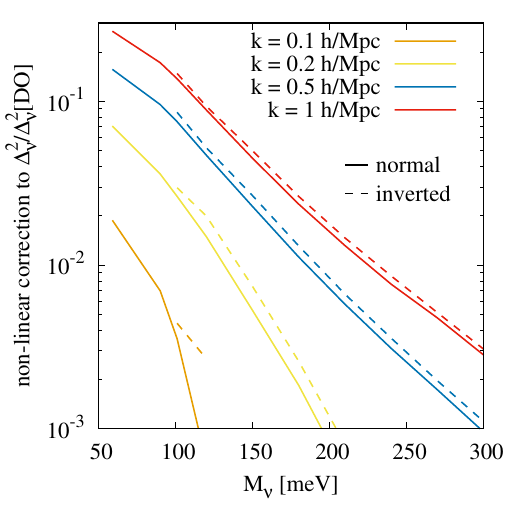}%
  \caption{
    Errors in the degenerate-ordering approximation in the neutrino power
    spectrum.  {\it Left}:~Total fractional error in the DO power spectrum
    $\Delta_\nu^2$[DO] as an approximation to $\Delta_\nu^2$[NO]~(solid) and
    $\Delta_\nu^2$[IO]~(dashed). 
    {\it Right}: Fractional non-linear contribution to the error in the DO
    approximation.  In all cases, we have used $\Nglq = N_\tau = 20$
    and $N_\mu = 10$.
    \label{f:error_in_DO}
  }
  \end{center}
\end{figure}

Figure~\ref{f:error_in_DO} shows how the DO approximation fares in the prediction of the neutrino power spectrum, when it is used at several $k$ to estimate the NO and IO power spectra over their respective mass ranges $\Mnu \geq 59$~meV and $\Mnu \geq 101$~meV.  As expected, the relative excess of the actual $\Delta_\nu^2$ over $\Delta_\nu^2$[DO] rises with decreasing $\Mnu$ and increasing $k$, with the degenerate ordering underestimating the power spectrum at $k=1~h/$Mpc and $\Mnu=59$~meV by a factor of more than thirty.  Figure~\ref{f:error_in_DO}~(Right) further shows that non-linear corrections alone, computed here  using Time-RG for the CB fluid and \FFTMII{} for the neutrinos, represent more than $10\%$ of this increase for $k \geq 0.5~h/$Mpc.  As constraints improve, we must be increasingly cautious about applying the degenerate-ordering approximation to studies of small-scale neutrino effects.

\subsection{Extension to axionic models}
\label{subsec:re1:extension_to_axionic_models}

\begin{figure}[t]
\begin{center}
  \includegraphics[width=0.99\textwidth]{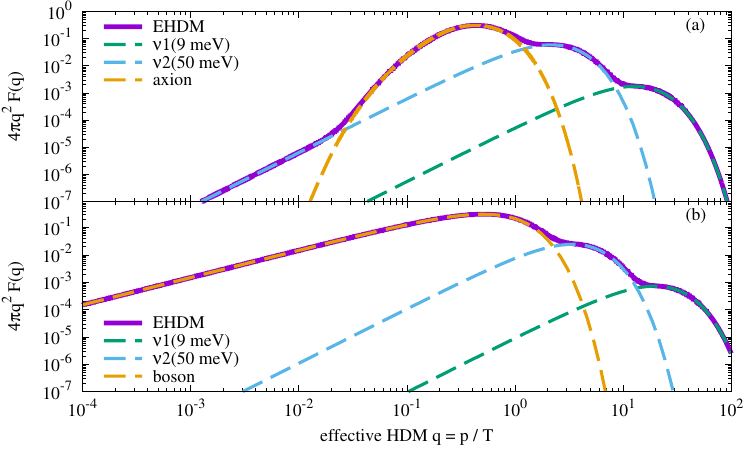}
  \caption{
    Effective distribution functions for two HDM models containing $\Mnu=59$~meV NO
    neutrinos, in addition to (a)~an axion whose distribution function was computed in
    Ref.~\cite{Notari:2022ffe}, or (b)~a generic thermal boson following a Bose-Einstein distribution.
    \label{f:F_ehdm_nuax_nuBE}
  }
  \end{center}
\end{figure}

We demonstrated in Sec.~\ref{subsec:eff:effective_hdm} that the EHDM formalism is not limited to neutrinos, but applies to any set of HDM species.  Here, we test its accuracy for models containing either axions or axion-like bosons, along with massive neutrinos.  We assume a cosmological constant as the dark energy, and we fix the following cosmological parameters:
\begin{equation}
  \Omo h^2 = 0.1424;\quad
  \Obo h^2 = 0.02242;\quad
  A_{\rm s} = 2.1\times 10^{-9};\quad
  n_{\rm s} = 0.966;\quad
  h = 0.6766.
  \label{e:cosmo_params_hyb}
\end{equation}
We assume minimal-mass NO neutrinos, $\Mnu=59$~meV, along with another  particle with mass $228$~meV, temperature $\Thdms=1.86$~K, and one of the following distribution functions:
\begin{enumerate}
\item[(a)] axionic, as computed in Ref.~\cite{Notari:2022ffe} for QCD
  axion production rates based upon pion-pion scattering data, and provided to us by the authors;
\item[(b)] bosonic, that is, the Bose-Einstein distribution function $F_{\rm BE}(q) = (2\pi)^{-3} (e^{q}-1)^{-1}$.
\end{enumerate}
The mass $228$~meV is chosen so that in the axionic case, the thermal axion population's contribution to $\Neff$ of $\Delta \Neff= 0.19$ remains slightly below observational bounds.  The bosonic case exceeds these bounds and is included for illustrative purposes.  Figure~\ref{f:F_ehdm_nuax_nuBE} shows the distribution functions of these two models.

As a high-accuracy reference calculation, we use a set of hybrid N-body simulations based upon the code of Ref.~\cite{Chen:2022dsv}, extended to EHDM models and using GLQ flows, as implemented in our companion paper, Ref.~\cite{Pierobon:2024XX}.  We conducted one hybrid simulation for each of the axionic and bosonic models above, and two more for models that include only standard neutrinos as the HDM, with NO masses totally $\Mnu=161$~meV and and $315$~meV, respectively.  All four simulation runs and their corresponding \FFTMII{} runs  use cosmological constant models with parameters given in \EQ{e:cosmo_params_hyb}, as well as $\Nglq = \Ntau=15$ and $\Nmu=10$.

\begin{figure}[t]
\begin{center}
  \includegraphics[width=0.99\textwidth]{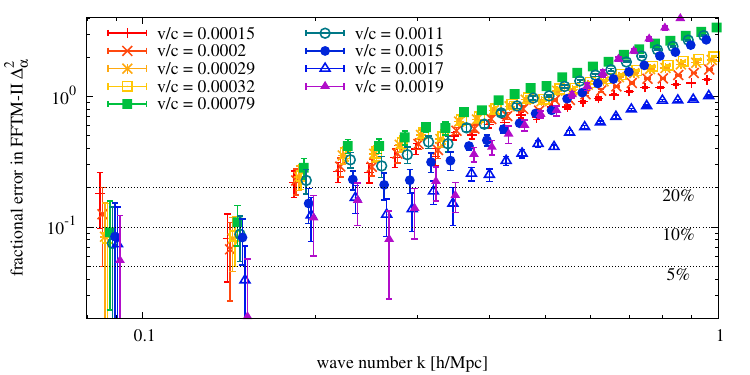}%

  \caption{
    Errors the in \FFTMII{} power spectra of the individual EHDM flows for a range of
    flow velocities relative to hybrid N-body simulation. The simulation results have been binned, with each data point representing the average over 20 points per bin and the error bars the standard deviation. The full set of data points are collated from four HDM models described in Sec.~\ref{subsec:re1:extension_to_axionic_models}, with other cosmological parameters fixed to \EQ{e:cosmo_params_hyb}.  For visual clarity, we have applied a horizontal offset of up to a few percent to the data points. 
    \label{f:err_vs_v}
  }
  \end{center}
\end{figure}

Reference~\cite{Chen:2022dsv} showed for standard neutrino models that \FFTM{} is accurate for flow velocities $v_\alpha/c \geq 0.0017-0.002$, that is, about $500$~km$/$sec$-600$~km$/$sec. The four models considered here together have a total of nine flows with $v_\alpha/c \leq 0.002$: $0.00032$ and $0.0017$ for the axionic model; $0.0002$ and $0.0011$ for the bosonic model; $0.00029$ and $0.0015$ for the lighter neutrino model; and $0.00015$, $0.00079$, and $0.0019$ for the heavier neutrino model.  Figure~\ref{f:err_vs_v} shows the fractional error in \FFTMII{} for each of these flows.  Evidently, low-$k$ error tends to decrease with rising flow velocity.  At larger wave numbers, $0.1~h/$Mpc$< k < 0.4~h/$Mpc, flows with $v/c \leq 0.001$ have similar errors, while errors in the higher-velocity flows decrease with increasing $v$.  Further, the two fastest flows shown, identified by filled and open triangles in the figure, have errors consistent with $\lesssim 10\%$ for $k \leq 0.2~h/$Mpc and $\lesssim 20\%$ up to $k \approx 0.35~h/$Mpc.  One of these two comes from the axion+neutrino model and the other one from a neutrino-only model.  Thus we see that the guideline of Ref.~\cite{Chen:2022dsv}, that perturbation theory is adequate for flows with $v/c$ larger than about $0.0017-0.002$, holds even for very different HDM species.

\begin{figure}[t]
\begin{center}
  \includegraphics[width=0.99\textwidth]{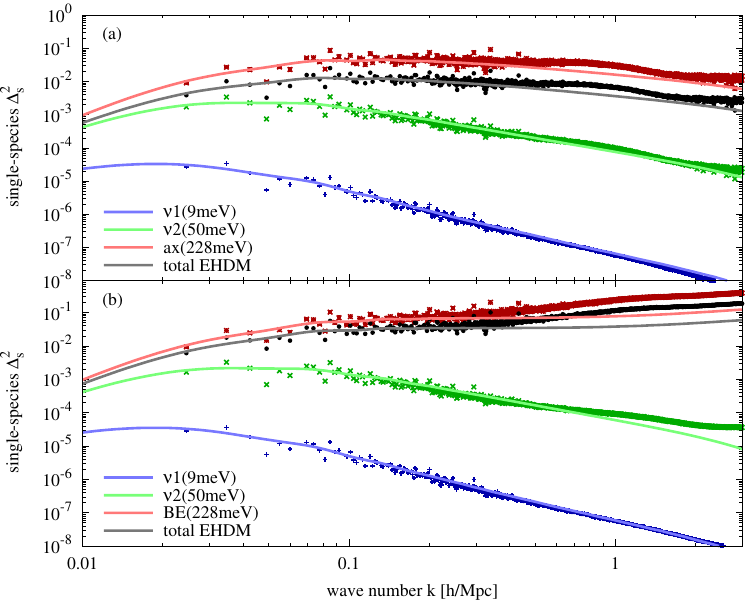}%

  \caption{
    Power spectra at $z=0$ for individual HDM species, determined from the GLQ
    flows as per \EQ{e:drho_species_glq}, for two HDM models.  Each model assumes a cosmological
    constant and the parameters of \EQ{e:cosmo_params_hyb}, with NO neutrino
    masses totally $\Mnu=59$~meV plus a non-standard HDM:
    (a)~axion+$\nu$ model, where the axion distribution was computed in Ref.~\cite{Notari:2022ffe}, and 
    (b)~boson+$\nu$ model, where the boson follows the relativistic Bose-Einstein distribution.
    \label{f:ax_BE_single_species}
  }
  \end{center}
\end{figure}

Figure~\ref{f:ax_BE_single_species} uses $\Nglq=15$ flows to compare \FFTMII{} and hybrid N-body power spectra for the axionic and bosonic models above.  In each case, the hybrid simulation converts into particles every flow with a velocity $v/c \leq 0.002$, i.e., the slowest two flows in each of these models, while the remaining flows are tracked using multi-fluid linear perturbation theory.  \FFTMII{} agrees closely with N-body power spectra in all cases up to $k \lesssim 0.4~h/$Mpc, and for the $9$~meV neutrinos to $k \approx 1~h/$Mpc.  We have thus demonstrated that $15$ flows are sufficient for predicting the clustering of the total EHDM as well as the component HDM species, even for species with very different distribution functions and an axion-to-neutrino mass ratio of $25$.

\section{Results II: Evading cosmological neutrino bounds}
\label{sec:re2}

\subsection{Non-standard neutrinos and clustering suppression}
\label{subsec:re2:non-standard_nu_and_clustering_suppression}

Cosmological upper bounds on $\Mnu$ are considerably stronger than those from laboratory experiments, such as $\Mnu < 2400$~meV from KATRIN~\cite{KATRIN:2021uub}, but also more dependent upon our assumptions that the neutrinos have only Standard Model interactions and an approximately Fermi-Dirac distribution function.  Moreover, the most stringent cosmological constraints rely upon  the assumption of $\Lambda$CDM cosmology.  Recently, Refs.~\cite{Oldengott:2019lke,Escudero:2022gez} and others have proposed models that allow massive neutrinos to evade cosmological bounds but whose mass remains measurable in ongoing and near-future laboratory $\beta$-decay end-point experiments including KATRIN.  Following Ref.~\cite{Escudero:2022gez}, we divide these models into two classes: ``skewed-$\nu$'' models, in which neutrinos' distribution function is skewed away from the relativistic Fermi-Dirac distribution by a momentum-dependent factor so as to increase their mean momentum; and ``cool-$\nu$'' models, which lower the neutrinos' temperature, hence their number density, in the early universe.

Reference~\cite{Alvey:2021sji} argues that cosmology chiefly constrains two properties of the relic neutrino background, its energy densities in the early and late universe, parameterized respectively as $\Neff$ and $\Ono$.  Thus, the goal of these alternative neutrino models is to preserve these two parameters while increasing $\Mnu$ into the range $600~{\rm meV} \lesssim \Mnu \lesssim 2400$~meV between the current laboratory bounds and the design sensitivity of ongoing experiments such as KATRIN.  The argument that late-time cosmology only constrains $\Ono$ follows from the standard result that neutrinos making up a matter fraction $\fnu = \Ono / \Omo$ cause an $\sim 8 \fnu$ fractional suppression of the matter power spectrum on scales much smaller than their free-streaming scale.  This is equivalent to a fractional suppression of $\delta_{\rm m}$ by $4 \fnu$ and of $\delcb$ by $3 \fnu$.  We begin this section by deriving this result in order to show its breadth as well as its limitations.

Applying \EQ{e:eom_delta_lin} to the CDM+baryon fluid (which has zero velocity) implies $-ak^2 \Phi = \Hc [a \Hc \delcb']'$.  Working in the Einstein-de Sitter model and considering $k$ sufficiently large that neutrino clustering can be neglected from \EQ{e:eom_Poisson}, we find $\delcb \propto a^{1-3\fnu/5}$ for small $f_\nu$, as compared with $\delcb \propto a$ in the massless-neutrino case.  However, these are only valid while neutrinos are non-relativistic, $a \gtrsim a_{\rm nr} = \bar p_{\nu,0} / m_\nu$, where the mean momentum $\bar p_{\nu,0} \approx 3.15 \Tnu$ for the Fermi-Dirac distribution; before this time, neutrino masses have a negligible impact upon $\delcb$.  Thus the late-time suppression factor of $\delcb$ is $\approx (a_{\rm nr}/a)^{3f_\nu/5}$.  This corresponds to a fractional suppression $1-(a_{\rm nr}/a)^{3f_\nu/5} \approx 3 f_\nu$ for $a=1$ and $\Ono h^2 \approx 0.003$, hence a suppression of $8 \fnu$ for the matter power spectrum.  The fractional suppression is weakly dependent upon $\Ono h^2$ and the scale factor; increasing $\Ono h^2$ to $0.01$ at $a=1$ leads to a suppression of $3.3 f_\nu$ in $\delta_{\rm cb}$ and hence $8.6 f_\nu$ in the matter power spectrum, while $\Ono h^2 = 0.003$ and $a=0.5$ gives a suppression of $2.6 f_\nu$ and $7.2 f_\nu$ in $\delta_{\rm cb}$ and the matter power spectrum, respectively.

They key point here is that, because $a_{\rm nr}$ depends on the ratio $p_{\nu,0} / m_\nu$,
increasing the neutrino momenta and masses by the same amount preserves this suppression due to neutrino free-streaming.  Thus skewed-$\nu$ models should have approximately the same small-scale power suppression as the corresponding standard-$\nu$ models characterized by $\Tnu = 1.9525$~K and Fermi-Dirac distribution functions.  However, cool-$\nu$ models, which seek to lower the neutrinos' temperature in order to increase their masses, will lead to different small-scale matter power spectra, requiring a modification to the arguments of  Ref.~\cite{Alvey:2021sji}.

We consider in the following the skewed-$\nu$ and cool-$\nu$ models in turn. Our benchmark observable is the CMB lensing potential power spectrum $C_L^{\phi\phi}$, which can probe matter clustering in the quasi-linear regime (corresponding to multipoles $1000 \lesssim L \lesssim 2000$) while remaining relatively free of systematic biases; see Ref.~\cite{Lewis:2006fu} for a review of CMB lensing and Ref.~\cite{McCarthy:2020dgq} regarding biases from baryonic effects.  As a criterion for discerning between the two models, we compare the differences between their $C_L^{\phi\phi}$ to the sensitivity forecast for a sample CMB Stage-4 survey~\cite{CMB-S4:2016ple}.   In order to predict $C_L^{\phi\phi}$, we combine the \hyphi{} code of Ref.~\cite{Upadhye:2023zgr} with the \FFTMII{} neutrino treatment of Sec.~\ref{sec:eff}.

\subsection{Skewed neutrino models}
\label{subsec:re2:skewed_nu_models}

Skewed-$\nu$ models were considered in, e.g., Refs.~\cite{Cuoco:2005qr,Oldengott:2019lke}.  Evidently from our discussion in Sec.~\ref{subsec:eff:effective_hdm}, the clustering of any collection of HDM species is determined by its velocity distribution.  Thus, increasing both their masses and their momenta so as to preserve the velocity distribution will have no impact upon the HDM clustering.  In other words, neutrinos could exceed cosmological bounds if their distribution function were skewed away from the relativistic Fermi-Dirac distribution and towards higher momenta.  The drawback of skewed-$\nu$ models is that no obvious mechanism for such a skew is known.

We parameterize the skewed neutrino distribution function as $F_{\rm sk}(q) = N_\sigma \sigma(q) \Ffd(q)$.  By demanding that the skewed-$\nu$ model reproduces the correct $\Neff$, that is, the  skewed-$\nu$ and standard-$\nu$ models must have the same energy densities at early (pre-Big Bang Nucleosynthesis) times:
\begin{equation}
  N_\sigma \int_0^\infty \frac{dq \, q^3 \sigma(q)}{e^q+1}
  =
  \int_0^\infty \frac{dq \, q^{3}}{e^q+1},
\end{equation}
we find a normalization factor
\begin{equation}
 N_\sigma = \frac{\Ifd_3}{\Isk_3},
\end{equation}
where
\begin{equation}
  \Ifd_n
  =
  \int_0^\infty \frac{dq \, q^{n}}{e^q+1},
  \quad{\rm and}\quad
  \Isk_n = \int_0^\infty \frac{dq \, q^{n} \sigma(q)}{e^q+1}.
\end{equation}
Here, $\Ifd_n$ is the standard Fermi-Dirac integral.  Next, we equate the models' late-time densities ${\bar \rho}_{\nu,0}  \propto \Ono$ in the non-relativistic approximation,
\begin{equation}
  \Mnuskew N_\sigma \int_0^\infty dq \frac{q^2 \sigma(q)}{e^q+1}
  =
  \Mnu  \int_0^\infty dq \frac{q^2}{e^q+1},
\end{equation}
which leads to
\begin{equation}
  \frac{\Mnu}{\Mnuskew}
  =
  \frac{\Isk_2}{\Ifd_2} \frac{\Ifd_3}{\Isk_3}.
  \label{e:massratio}
\end{equation}
On the other hand, the mean momentum is readily seen to be $a^{-1} \Tnu \Ifd_3 / \Ifd_2 \approx 3.15 \Tnu/a$ for standard neutrinos and $a^{-1} \Tnu \Isk_{3}/\Isk_{2}$ for skewed neutrinos.  Then, by comparison with \EQ{e:massratio}, we see immediately that the ratios of mean momentum to neutrino mass are the same in both models.  Thus, skewed-$\nu$ models must also cause a small-scale fractional suppression of $\sim 8 f_\nu$ in the late-time matter power spectrum.

As a phenomenological example, we consider a neutrino distribution function $F_{\rm sk}(q) = N_{\nskew} q^{\nskew} \Ffd(q)$, where $N_{\nskew}$ is the normalization constant, and a larger $\nskew > 0$ implies a larger mean momentum.  We assume that the normalization is fixed in the early universe so as to preserve $\Neff=3.044$.  This means that increasing $\nskew$ reduces the neutrinos' number density, such that fixing $\Ono$ would require a higher $\Mnu$.  Alternatively, we could fix $\Mnuskew=600$~meV while increasing $\nskew$: doing so would allow the skewed-$\nu$ model to mimic a standard neutrino model with smaller $\Mnu$; indeed, we find that a skewed model with $\nskew=13$ ($\nskew=28$) has the same $\Ono$ as standard neutrinos with $\Mnu=118$~meV ($61$~meV).  Although this example does not exactly preserve the velocity distribution function, we will see that it is indistinguishably close to the standard neutrino case for near-future cosmological surveys.

\begin{figure}[t]
\begin{center}
  \includegraphics[width=0.99\textwidth]{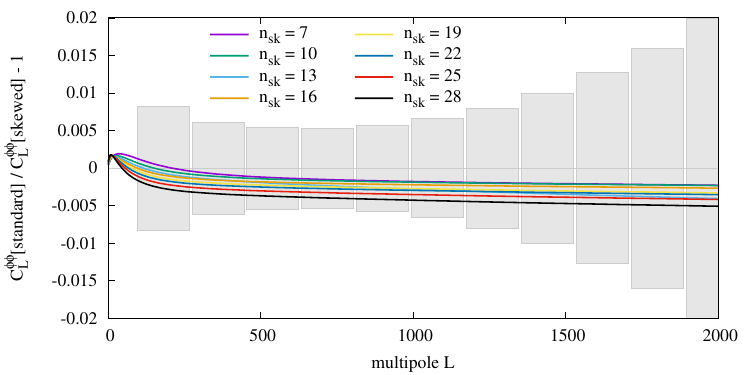}%
  \caption{
    Fractional difference in the CMB lensing potential power spectra between
    skewed-$\nu$ models with distribution functions
    $\propto  p^{\nskew} \Ffd(p)$ and $\Mnuskew=600$~meV, 
    and standard-$\nu$ models described by a relativistic Fermi-Dirac distribution $\Ffd(p)$. The skewed-$\nu$ distribution functions have been normalized to ensure $\Neff=3.044$.  For each $\nskew$, the corresponding standard-$\nu$ mass $\Mnu$ has been chosen such that it gives an energy density $\Ono$ matching its skewed-$\nu$ counterpart.  That is, $\Ono$ decreases with rising $\nskew$. The shaded error bands correspond to the forecasted CMB Stage-4 sensitivities taken from Ref.~\cite{CMB-S4:2016ple}.
    \label{f:CLpp_OBKSS}
  }
  \end{center}
\end{figure}

Figure~\ref{f:CLpp_OBKSS} shows our results with fixed $\Mnuskew=600$~meV.  Each skewed-$\nu$ model is compared with its standard-$\nu$ counterpart, with matching $\Neff$ and $\Ono$ .  The fractional difference between the CMB lensing potential power spectra in each pair fits comfortably within the binned error bands forecast for the proposed CMB Stage-4 survey in the CMB-S4 Science Book~\cite{CMB-S4:2016ple}.  Thus, as expected, raising the neutrino masses and momenta in tandem does not appreciably affect neutrinos' clustering properties.  If a theoretically-suitable, experimentally-falsifiable mechanism could be found for introducing such a distortion to the neutrinos' distribution function, then skewed neutrino models would be viable ways to evade cosmological neutrino mass bounds.

\subsection{Cool-neutrino models}
\label{subsec:re2:cool_nu_models}

Contrary to skewed-$\nu$ models, cool-$\nu$ models increase the neutrinos' masses while decreasing their mean momenta, both of which have the effect of reducing the neutrinos' velocities, thereby amplifying both their small-scale power suppression and their non-linear clustering while shifting their free-streaming wave numbers $\kfs$ to larger values.  We consider here the cool-$\nu$ model of Ref.~\cite{Escudero:2022gez}, where neutrinos thermalize 
with $N_\chi$ massless sterile particles $\chi$ via a massive mediator at a time after weak decoupling.
The resulting neutrinos have a relativistic Fermi-Dirac distribution function.  However, equipartition amongst the thermalized species leads to a neutrino temperature that is lower by a factor of $(1+2N_\chi/3)^{-1/3}$ compared with Standard Model neutrinos; the remainder of $\Neff$ is made up by the massless particles.  These cooler neutrinos have a correspondingly reduced number density, meaning that their masses must be larger than those of standard neutrinos for the same $\Ono$ value.

For example, a factor-of-two cooling in the neutrinos, corresponding to $N_\chi \approx 11$, implies an eightfold reduction in their number density, allowing for a corresponding eightfold increase in the sum of their masses, $\Mnucool$.  Since the neutrino free-streaming wave number $\kfs \propto m / T$, both the reduced temperature and the increased mass raise $\kfs$, pushing the neutrino suppression to smaller scales.  In this particular example of $N_\chi \approx 11$, $\kfs$ increases by a factor of sixteen.

We estimated in Sec.~\ref{subsec:re2:non-standard_nu_and_clustering_suppression} that the late-time fractional suppression in $\delcb$ by massive neutrinos is $\approx (a_{\rm nr}/a)^{3f_\nu/5}$.  Preserving $f_\nu$ while reducing $\Tnu$ and increasing $m_\nu$ will therefore reduce $a_{\rm nr}$, increasing the suppression.  Thus a cool-$\nu$ model with the same $\Neff$ and $\Ono$ as a standard-temperature model will nevertheless have a smaller free-streaming scale and a greater suppression of the linear $\delcb$ and hence the linear matter power spectrum.  Continuing with our $N_\chi \approx 11$ example of a halving of $\Tnu$ and an eightfold increase in $\Mnucool$ relative to standard-temperature models, we find for $\Ono h^2 \approx 0.003$ and $a=1$ that $\delcb$ and the matter power spectrum are suppressed at small scales by $\approx 4.5\fnu$ and $\approx 11 \fnu$, respectively.

\begin{figure}[t]
\begin{center}
  \includegraphics[width=0.99\textwidth]{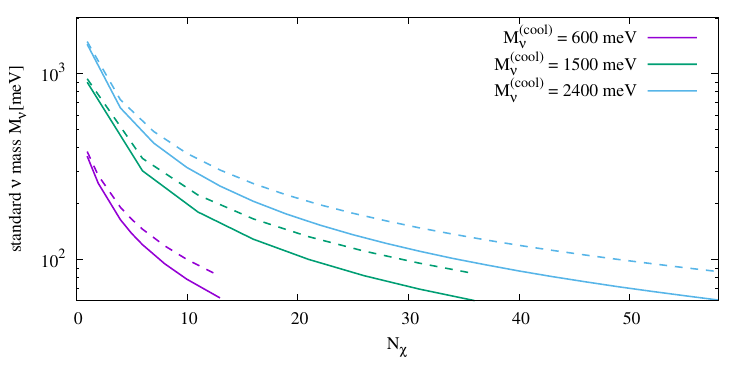}%

   \caption{Neutrino mass sums $\Mnu$ in standard-$\nu$ models required to match the late-time energy density
     $\Ono$ (solid) and the small-scale matter power suppression (dashed) of cool-$\nu$ models with $\Mnucool=600$~meV (purple), $1500$~meV (green), and $2400$~meV (blue).  The maximum $N_\chi$ shown
     is consistent with the equal-$\Ono$ mass sum $\Mnu$ being at least $60$~meV.
     \label{f:cool-nu_eq_supp_M} 
   }
   \end{center}
\end{figure}

Since no single $\Mnucool$ value in a cool-$\nu$ model can simultaneously match the late-time density parameter $\Ono$ and the small-scale matter power spectrum of a standard-$\nu$ model, 
we next ask which of these two choices of late-time phenomenology is the better approximation to cosmological constraints.  We compare cool-$\nu$ models with fixed $\Mnucool$ against two different standard-$\nu$ analogs: one with an equal $\Ono$, and another with an equal small-scale suppression.  All models have the same $\Neff$ by design.  In the equal-$\Ono$ case, assuming $N_\chi$ massless sterile particles, the standard-$\nu$ model has $\Mnu = \Mnucool / (1 + 2N_\chi/3)$. 

In the equal-suppression case, since our suppression formula above is  approximate, we match the $z=0$ linear matter power suppression computed using \classcode{} at $k=10~h/$Mpc to a fractional precision $< 10^{-5}$ by adjusting $\Mnu$.  We assume $\nu\Lambda$CDM models with parameters
\begin{equation}
  \Omo h^2 = 0.14;\quad
  \Obo h^2 = 0.022;\quad
  A_{\rm s} = 2.2\times 10^{-9};\quad
  n_{\rm s} = 0.96;\quad
  h=0.67.
  \label{e:cosmo_params_low-om}
\end{equation}
We further assume NO masses in the standard-$\nu$ case.  Since we are interested in $\Mnucool \geq 600$~meV, we make the DO mass approximation for cool-$\nu$ models.  Figure~\ref{f:cool-nu_eq_supp_M} shows equal-density and equal-suppression masses as functions of $N_\chi$ for $600$~meV$\leq \Mnucool \leq 2400$~meV.

\begin{figure}[t]
\begin{center}
  \includegraphics[width=0.99\textwidth]{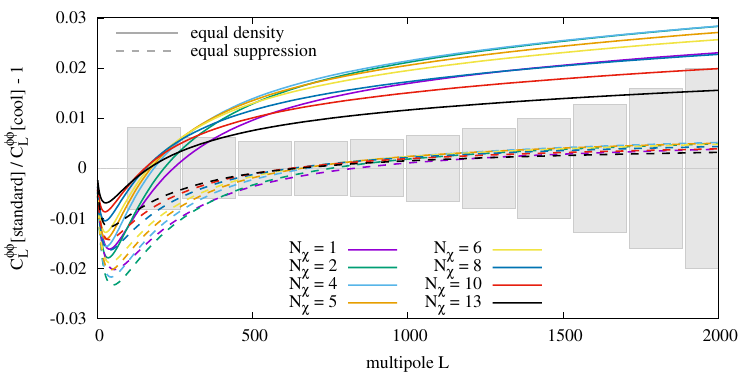}%

  \caption{
    Fractional differences in the CMB lensing potential power spectrum
    $C_L^{\phi\phi}$ between standard-$\nu$ and cool-$\nu$ models with the
    cosmological parameters of \EQ{e:cosmo_params_low-om}.  We fix
    $\Mnucool=600$~meV, while varying the number $N_\chi$ of sterile species. 
    Solid and dashed lines represent, respectively, standard-$\nu$ models matched to the 
    same $\Ono$ and the same small-scale suppression as the
    corresponding cool-$\nu$ model. The shaded error bands denote the forecasted
    CMB Stage-4 sensitivities taken from Ref.~\cite{CMB-S4:2016ple}.
    \label{f:CLpp_cool-nu}
  }
  \end{center}
\end{figure}

Figure~\ref{f:CLpp_cool-nu} compares the CMB lensing potential power spectra of equal-density and equal-suppression standard-$\nu$ models, to the corresponding cool-$\nu$ model of a fixed $\Mnucool=600$~meV and various choices of $N_\chi \leq 13$.  Increasing $N_\chi$ above $13$ will imply equal-density $\Mnu$ below the lower bound from neutrino oscillation experiments, so we exclude these from consideration.  For $N_\chi=6$ ($N_\chi=13$), the equal-density standard-$\nu$ model has $\Mnu=120$~meV ($\Mnu=62$~meV), while the corresponding equal-suppression masses are $\approx 21\%$ ($\approx 31\%$) larger.  Evidently, equal-density standard-$\nu$ models will be distinguishable from the corresponding cool-$\nu$ models to a high significance using CMB Stage-4 data, even for the smallest masses $\Mnucool=600$~meV accessible to ongoing terrestrial experiments.  Thus, from a phenomenological viewpoint, the statement that late-time cosmological observable depends only on the HDM density $\Ono$ is not correct, as demonstrated here by the cool-$\nu$ models.

On the other hand, as shown in Fig.~\ref{f:CLpp_cool-nu}, equal-suppression standard-$\nu$ and cool-$\nu$ models exhibit much the same CMB lensing potential power spectrum over a large range of multipoles $L$.  In the case of 
$N_\chi \geq 10$, the equal-suppression standard-$\nu$ model lies within the error bars over the entire $L$ range considered. Thus the HDM property actually constrained by the late-time cosmological data is closer to the small-scale matter power spectrum suppression than the background HDM density.  While this argument leaves intact the most important point made by Ref.~\cite{Escudero:2022gez}, namely, the ability of cool-$\nu$ models to evade neutrino mass constraints from observational cosmology, it also suggests new approaches by which cool-$\nu$ models may be excluded in the future.  The key point is that the precise $\Mnu$ characterizing an equal-suppression model depends upon the observable as well as the scale factor.  Specifically:
\begin{enumerate}
\item The linear suppression factor is weakly dependent upon the scale factor,
  as noted in Sec.~\ref{subsec:re2:non-standard_nu_and_clustering_suppression}, with a threefold change in the scale factor leading to a
  $\approx 10\%$ change in the linear suppression fraction.
\item Cool-$\nu$ models, by virtue of having larger masses and lower
  temperatures than equal-suppression standard-$\nu$ models, have larger
  non-linear HDM clustering, itself depending rapidly upon the scale factor.
  \label{i:NL_HDM_clustering}
\item The cool-$\nu$ and equal-suppression standard-$\nu$ models have
  different $\fnu$, so the ratios of small-scale linear matter-to-cb suppression
  factors $\lim_{k\rightarrow\infty} P_{\rm m}(k)/P_{\rm cb}(k) = (1-\fnu)^2$
  are also different.
\end{enumerate}
CMB lensing and tomographic shear surveys probe the matter power spectrum at different redshifts, while galaxy clustering surveys trace the cb power spectrum.  Thus a combination of all three may be able to eliminate or severely constrain cool-$\nu$ models designed to evade cosmological bounds.  Furthermore, a large-volume survey may be able to constrain the difference between the free-streaming scales of cool-$\nu$ and standard-$\nu$ models.  We leave forecasts of such joint constraints to future work.

\begin{figure}[t]
\begin{center}
  \includegraphics[width=0.49\textwidth]{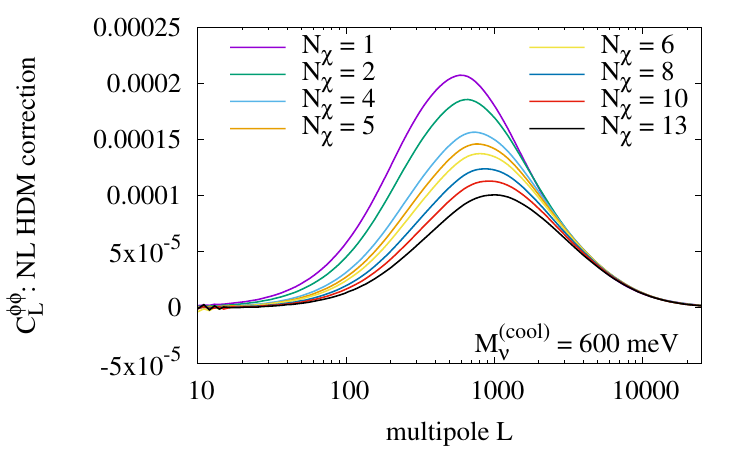}
  \includegraphics[width=0.49\textwidth]{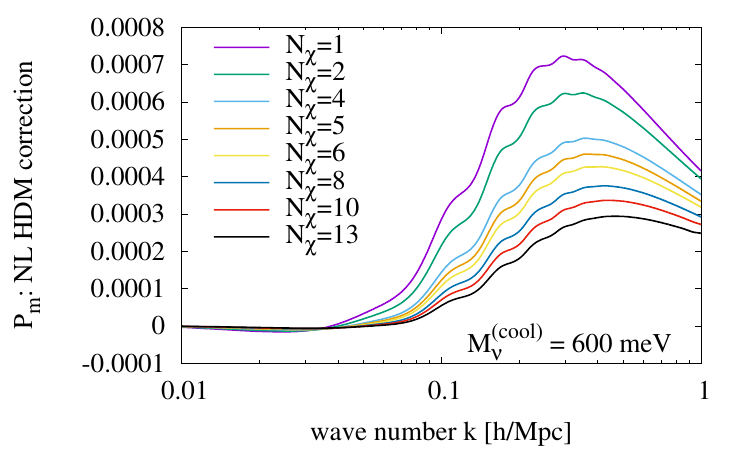}%

  \includegraphics[width=0.49\textwidth]{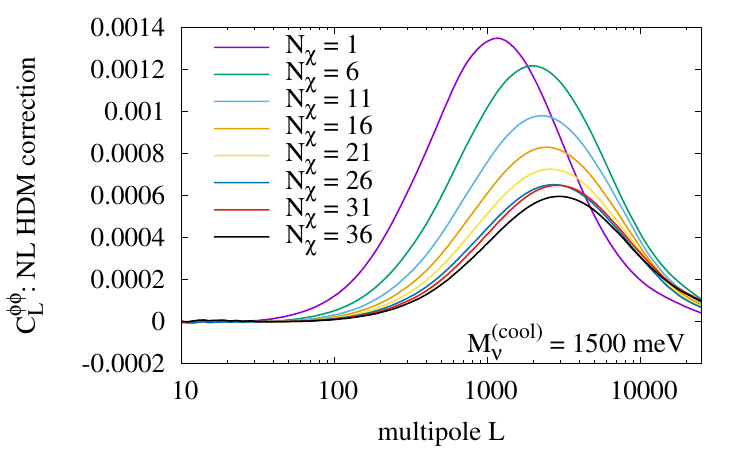}
  \includegraphics[width=0.49\textwidth]{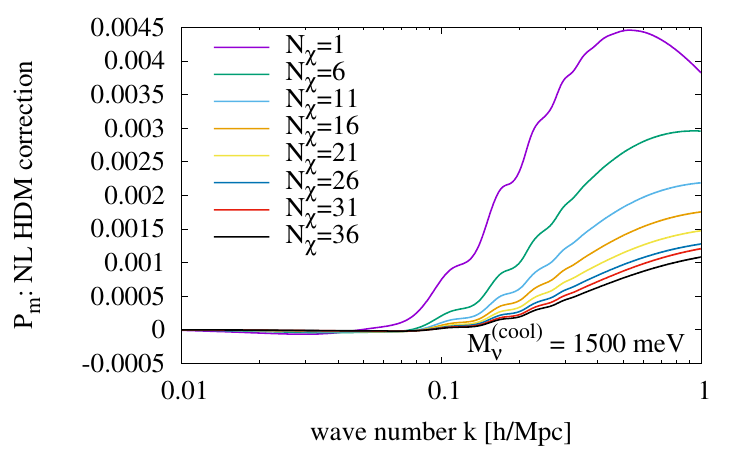}%

  \includegraphics[width=0.49\textwidth]{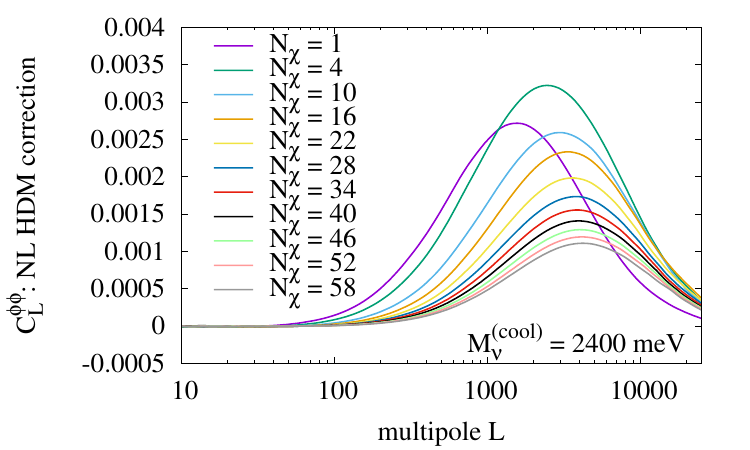}
  \includegraphics[width=0.49\textwidth]{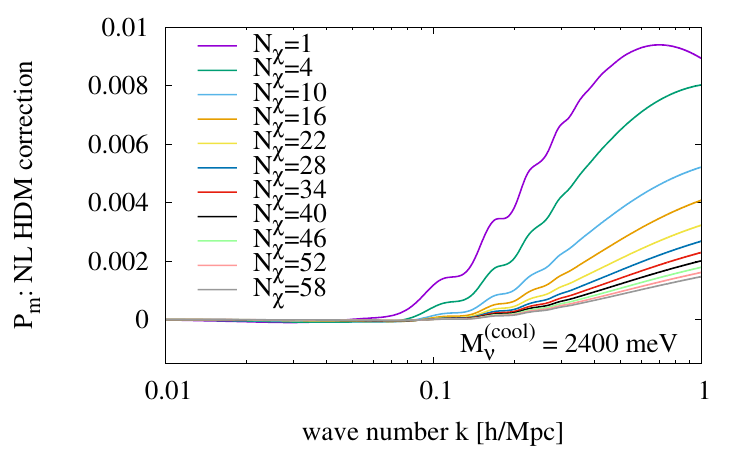}%

  \caption{
    Fractional corrections in cosmolgoical observables from neutrino non-linearity in cool-$\nu$ models. {\it Left}: Corrections to the CMB lensing potential power spectrum
    $C_L^{\phi\phi}$ for $\Mnucool=600$~meV (top), $1500$~meV (middle), and
    $2400$~meV (bottom) for various choices of $N_\chi$.
    {\it Right}: Corrections to the $z=0$ matter power spectrum $P_{\rm m}(k)$ for the same set of $\Mnucool$ and $N_\chi$.  The corresponding masses $M_\nu$ in standard-$\nu$ models, matching either the late-time density $\Ono$ or small-scale suppression of the cool-$\nu$ models,  are shown in Fig.~\ref{f:cool-nu_eq_supp_M}.    
    \label{f:cool-nu_NL}
  }
\end{center}
\end{figure}

Lastly, we elaborate upon item~\ref{i:NL_HDM_clustering} in the list above, the non-linear contribution to HDM clustering.  Figure~\ref{f:cool-nu_NL} quantifies the fractional contribution of \FFTMII{} non-linear neutrino corrections to the matter power spectrum and the CMB lensing potential power spectrum.  We consider cool-$\nu$ models with $\Mnucool$ of $600$~meV, $1500$~meV, and $2400$~meV, spanning the range of interest between the current terrestrial constraints of KATRIN~\cite{KATRIN:2021uub} and its ultimate design sensitivity.  
Adopting a conservative cosmological upper bound on $\Mnu$ of about $380$~meV~\cite{Planck:2018vyg},%
\footnote{Such a conservative bound could come from the simultaneous variation of the dark energy equation of state, its derivative, and the galaxy bias or Halo Occupation Distribution parameters, in the data analysis. See, e.g., Ref.~\cite{Upadhye:2017hdl} for details.}
we see that in order for $\Mnucool=2400$~meV to stay within observational constraints, a $N_\chi$ of at least $10$ would be required.  In this case, the bottom panels of Fig.~\ref{f:cool-nu_NL} show that the non-linear HDM correction to $C_L^{\phi\phi}$ is $0.16\%$ at $L=1000$ and $0.24\%$ at $L=2000$, while the small-scale $z=0$ matter power correction is $0.52\%$.  While even the $C_L^{\phi\phi}$ correction is a sizable fraction of the error bars and cannot be neglected, the rapid late-time increase of this correction could prove useful constraining it.


\section{Conclusions}
\label{sec:con}

We have developed and derived a procedure for representing an arbitrary collection of HDM species, of any masses, temperatures, and distribution functions, from the relativistic to the non-relativistic regime, using a single EHDM species with an appropriately-chosen distribution function.  As this method follows directly from the full collisionless Boltzmann equation, it makes no assumption about the nature of the inhomogeneities in any distribution function and is equally applicable to linear and non-linear perturbation theory, as well as to N-body simulations.  In this work, we have implemented this EHDM method in both linear (\MuFLR{}) and non-linear (\FFTMII{}) multi-flow perturbation theories, which discretize the EHDM distribution into $\sim 10$ uniform-momentum flows.  Furthermore, since terrestrial experiments are typically sensitive to only a single HDM species such as the electron neutrino, we have shown how an appropriate linear combination of these flows allows us to recover the power spectrum of each component HDM species that makes up the EHDM.

As cosmology enters the HDM era, perturbation theory is emerging as an indispensable tool, both for testing approximations within the standard neutrino picture and for exploring models well beyond it.  Within the standard picture, we have shown that a more efficient choice of flow momenta improves the small-scale accuracy of \FFTMII{} at low $\Mnu$ by more than a factor of two relative to its predecessor, and we have quantified the differences among the normal, inverted, and degenerate neutrino mass orderings for a range of $\Mnu$.  Beyond standard neutrinos, we have considered mixed-HDM models that incorporate either a thermal QCD axion or a generic thermal boson in addition to a minimally-massive neutrino sector.  We have also studied two attempts at evading cosmological neutrino bounds, either by skewing the neutrinos' distribution function or by reducing their temperature, and we have shown that the latter of these modifies the linear and non-linear clustering of neutrinos in a manner that could allow it to be distinguished from standard-temperature neutrinos using upcoming data from CMB Stage-4 experiments.  In doing so, we have demonstrated \FFTMII{} to be an invaluable tool for studying light neutrinos as well as rapidly exploring the non-standard HDM parameter space.

\subsection*{Acknowledgments}
We are grateful to J.~Kwan, Y.~Li, I.~G.~McCarthy, A.~Notari, and T.~Tan for insightful conversations.  This research was undertaken using the HPC systems \emph{Gadi} from the National Computational Infrastructure (NCI) supported by the Australian Government, and \emph{Katana} at UNSW Sydney. MM acknowledges support from the DFG grant LE 3742/8-1.  GP and Y$^3$W are supported by the Australian Government through the Australian Research Council’s Future Fellowship (FT180100031) and Discovery Project (DP240103130) schemes.


\bibliographystyle{bibi}
\bibliography{FlowsForTheMassesII}
\end{document}